\documentclass[fleqn,usenatbib]{mnras}

\usepackage{newtxtext,newtxmath}
\usepackage{mathptmx}

\usepackage[T1]{fontenc}

\DeclareRobustCommand{\VAN}[3]{#2}
\let\VANthebibliography\thebibliography
\def\thebibliography{\DeclareRobustCommand{\VAN}[3]{##3}\VANthebibliography}

\usepackage[table]{xcolor}
\usepackage{graphicx}	
\usepackage{comment}
\usepackage{hyperref}
\usepackage{caption}
\usepackage{subcaption}
\usepackage{multirow}
\usepackage{arydshln}
\usepackage{amsmath}	
\usepackage{lmodern}

\usepackage{array, booktabs, makecell}
\usepackage{siunitx}
\usepackage[version=4]{mhchem}

\title[Activity-dependence of quiet-Sun RVs]{Measuring the Sun’s radial velocity variability due to supergranulation over a magnetic cycle.}

\author[N. K. O'Sullivan et al]{Niamh K. O'Sullivan$^{1}$,\thanks{E-mail: niamh.osullivan@physics.ox.ac.uk}
Suzanne Aigrain$^{1}$,
Michael Cretignier$^{1}$,
Ben Lakeland $^{2}$, 
Baptiste Klein$^{1}$,  \and
Xavier Dumusque$^{3}$,
Nadège Meunier$^{4}$, 
Sophia Sulis$^{5}$,
Megan Bedell$^{6}$
Annelies Mortier$^{2}$, \and
Andrew Collier Cameron$^{7}$, and 
Heather M. Cegla $^{8}$$^{,}$$^{9}$
\\
$^{1}$Denys Wilkinson Building, Department of Physics, University of Oxford, OX1 3RH, UK\\
$^{2}$School of Physics and Astronomy, University of Birmingham, Edgbaston, Birmingham, B15 2TT\\
$^{3}$Observatoire astronomique de l’Université de Genève, 51 ch. des Maillettes, 1290 Versoix, Switzerland\\
$^{4}$Univ. Grenoble Alpes, CNRS, IPAG, F-38000 Grenoble, France\\
$^{5}$Aix Marseille Univ, CNRS, CNES, LAM, Marseille, France\\
$^{6}$Center for Computational Astrophysics, Flatiron Institute, 162, 5th Avenue, New York, NY 10010, USA \\
$^{7}$Centre for Exoplanet Science / SUPA, School of Physics \& Astronomy, University of St Andrews, North Haugh ST ANDREWS, Fife KY16 9SS, UK  \\
$^{8}$Centre for Exoplanets and Habitability, University of Warwick, Coventry CV4 7AL, UK \\
$^{9}$Department of Physics, University of Warwick, Coventry CV4 7AL, UK
}

\date{Accepted XXX. Received YYY; in original form ZZZ}

\pubyear{2024}

\begin{document}
\label{firstpage}
\pagerange{\pageref{firstpage}--\pageref{lastpage}}
\maketitle

\begin{abstract}
In recent years supergranulation has emerged as one of the biggest challenges for the detection of Earth-twins in radial velocity planet searches. We used eight years of Sun-as-a-star radial velocity observations from HARPS-N to measure the quiet-Sun's granulation and supergranulation properties of most of its 11-year activity cycle, after correcting for the effects of magnetically active regions using two independent methods. In both cases, we observe a clear, order of magnitude variation in the time-scale of the supergranulation component, which is largest at activity minimum and is strongly anti-correlated with the relative Sunspot number.  We also explored a range of observational strategies which could be employed to characterise supergranulation in stars other than the Sun, showing that a comparatively long observing campaign of at least 23 nights is required, but that up to 10 stars can be monitored simultaneously in the process. We conclude by discussing plausible explanations for the "supergranulation" cycle.
\end{abstract}




\section{Introduction}
\label{sec:intro}

Over the next decade, extremely precise radial velocity (EPRV) instruments will in principle reach the precision needed to detect Earth-like planets. Since the detection of the first exoplanet around a Sun-like star in 1995 \citep{1995Natur.378..355M}, over 5800 exoplanets have been confirmed. Most were detected by space-based transit-search missions such as \emph{Kepler} \citep{2010Sci...327..977B} and the Transiting Exoplanet Survey Satellite (TESS) \citep{2014SPIE.9143E..20R}. The PLATO mission \citep{2024arXiv240605447R}, due to launch in 2026, will combine a \emph{Kepler}-like collecting area with an extremely wide field-of-view in order to find Earth-like planets around nearby, Sun-like stars, suitable for further characterisation. Radial velocity (RV) follow-up is key to confirm transiting planet candidates and to determine the planet's mass. At the same time, blind EPRV searches employing a very intensive monitoring strategy will aim to detect non-transiting Earth analogues around the same stars \citep{2016SPIE.9908E..6FT, 2021AJ....161..130G, 2018MNRAS.479.2968H}. The key challenge in both cases is the intrinsic variability of the host star, which typically overwhelms the planetary signals of interest, and can also mimic them in some cases \citep[e.g.][]{2016MNRAS.456L...6R,2021arXiv210714291C,2024CRPhy..24S.140M}.

For the past 15 years, precise RV spectrographs such as  HARPS \citep{2003Msngr.114...20M},  and HARPS-N \citep{2012SPIE.8446E..1VC}  have made the detection of small planets, with RV semi-amplitudes down to around 1\,m/s, a matter of routine \citep{Cretignier(2023), 2024MNRAS.534.2410D, 2025A&A...693A.297N}, but have struggled to reach below this "floor". Their sensitivity is limited in part by night-to-night calibration errors of order $0.5$\,m/s \citep{2021A&A...648A.103D}. This motivated the development of a new generation of ultra-stable RV spectrographs equipped with Laser Frequency Combs (LFCs) for wavelength calibration, including EXPRES \citep{2016SPIE.9908E..6TJ}, KPF \citep{2016SPIE.9908E..70G}, NEID \citep{2018SPIE10702E..71S},  and ESPRESSO \citep{2021A&A...645A..96P}, which should reach instrument stability of $\sim 20$\,cm/s. However, even these state-of-the-art instruments are limited by intrinsic stellar variability, which introduces RV variations up to several m/s on a wide range of timescales. Next-generation RV surveys are thus planning to adopt very intensive monitoring strategies, to characterise these variations well enough to disentangle them from planetary signals from Earth analogues (with semi-amplitudes of order 10\,cm/s and periods of several hundred days).

An example of such a survey is the Terra Hunting Experiment (THE) \citep{2016SPIE.9908E..6FT, 2018MNRAS.479.2968H}, which will use the HARPS3 spectrograph, an upgraded copy of HARPS and HARPS-N that will be installed on the Isaac Newton Telescope (INT) on La Palma. HARPS3 will be equipped with an LFC and will observe bright stars at high Signal-to-Noise Ratio (SNR) to reach a precision of $\sim 30$\,cm/s per epoch, but what sets THE apart is that it will focus on a comparatively small number of target stars (a few dozen), which will be monitored very intensively for the full, 10-year duration of the survey. To enable this mode of operations, the INT has been refurbished and is fully robotic. While the baseline observing strategy is to observe each of the target stars once per night (subject to visibility constraints), this can be refined further; the optimal strategy depends on the detailed variability properties of the host stars and the strategies used to mitigate that variability \citet{2019Geosc...9..114C}.

\subsection{Supergranulation}

The interplay of convection and magnetic fields in the surface layers of Sun-like stars gives rise to complex structures in their photospheres, which induce RV variability on a wide range of time-scales. Active regions (regions of enhanced surface magnetic flux, containing spots, faculae and plages), give rise to RV variations with amplitudes of several m/s \citep{1997ApJ...485..319S, 2007A&A...473..983D, 2010A&A...512A..39M, 2011arXiv1107.5325L, 2012A&A...541A...9G} on timescales ranging from weeks (associated with the star's rotation rate and the intrinsic evolution of the active regions) to years (associated with cyclic behaviour of the dynamo that powers the star's large-scale magnetic field). Historically, this stellar activity signal has been the main factor limiting the detection of small-amplitude, long-period planets in RVs. However, the past 15 years have seen major progress in mitigating activity signals in RV. At the simplest level, decorrelating the RVs against spectroscopic activity indicators (such as $\log R'_{\rm HK}$ \& H$\alpha$) can significantly reduce the activity signals  \citep[see][for examples]{2009A&A...495..959B, 2011A&A...535A..55D, 2013ASPC..472..137F, 2021AJ....161..272H}, though such a simplistic approach is generally insufficient to capture their full complexity. More flexible models such as Gaussian Process (GP) regression \citep{2023ARA&A..61..329A} have become increasingly popular to model activity signals in RVs \citep{2014MNRAS.443.2517H}, and have proved particularly effective when applied to RVs and activity indicator time-series simultaneously \citep{2012MNRAS.419.3147A, 2015MNRAS.452.2269R, 2022MNRAS.509..866B, 2022A&A...659A.182D}. In recent years, new methods have been proposed to model line-shape changes induced by active regions directly in time-series of cross-correlation functions (CCFs) \citep{2021MNRAS.505.1699C,2022MNRAS.512.5067K, 2022AJ....164...49D}, and even spectra and/or line-by-line (LBL) RVs \citep{2017arXiv171101318J, 2018A&A...620A..47D, 2020MNRAS.492.3960R, 2022A&A...659A..68C, 2022MNRAS.513.5328L, Cretignier(2023), 2024A&A...687A.281Z}. In the best cases, these methods enable the correction of activity signals down to the sub-m/s level, allowing for the detection of low-mass exoplanets \citep[e.g.][]{2022A&A...658A.115F}.

At the opposite end of the frequency spectrum, stellar oscillations, caused by trapped sound waves in the stellar atmosphere \citep{1994ApJ...424..466G, 1995A&A...293...87K}, induce RV variations on timescales of a few minutes and with amplitudes of around 1\,m/s for Sun-like stars. These are typically mitigated by using exposure times that are similar to or longer than the oscillation periods, so that the RV signal is averaged out \citep{2019AJ....157..163C}. 

On slightly longer timescales, granulation is caused by convective upflows resulting in an overall convective blueshift \citep{1995ApJ...444L.119R, 2008A&A...490.1143L, 1991A&A...248..245R}. On the Sun, granulation flows can reach vertical velocities of several hundred m/s, and vary on timescales of a few 10s of minutes within individual granules, which themselves have a typical size of 1 Mm \citep{2021arXiv210406072M}. However, the disk-integrated RV signal, which is averaged over $1\,000\,000$'s (for a Sun-like star) of granules, is observed to be around 0.4\,m/s for the Sun \citep{1994MNRAS.269..529E, 1999ASPC..173..297P, 2020A&A...635A.146S}.

\begin{figure}
    \centering
    \includegraphics[width=0.44\textwidth]{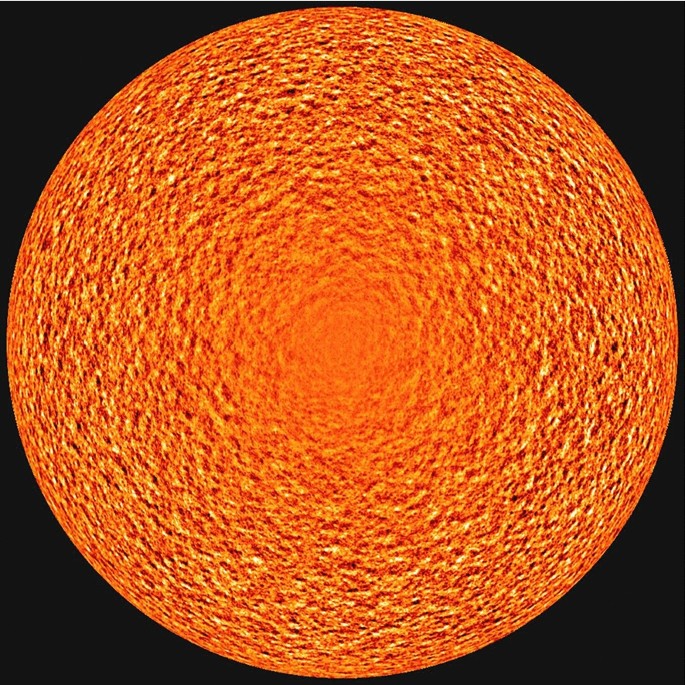}
    \caption{ Dopplergram of the sun showing the supergranulation pattern. Due to the horizontal nature of supergranulation, the granules are more visible towards the limb of the Sun. (image credit: SOHO/MDI/ESA)}
    \label{fig:supergranulation}
\end{figure}

While granulation mainly gives rise to vertical flows in the photosphere, supergranulation (SG), first discovered 60 years ago \citep{1954MNRAS.114...17H, 1956MNRAS.116...38H} is a predominantly horizontal flow. It was first discovered as a horizontal velocity field super-imposed on the Sun's mean solar rotation field near its equator, and confirmed as a disk-wide phenomenon by \citet{1962ApJ...135..474L} using Doppler images. Supergranulation is a velocity perturbation below of the granulation field, with each supergranule in the Sun measuring 30--35\,Mm across \citep{1994SoPh..154....1N, 2010A&A...512A...4R, 2017A&A...599A..69R}, which share a predominantly horizontal motion, with horizontal speeds of 200--400\,m/s compared to vertical speeds of 20--30\,m/s \citep{2018LRSP...15....6R}. The predominantly horizontal nature of the supergranulation flow is evident in SDO Dopplergram images after subtraction of the rotational motion of the Sun and the mean convective blueshift, where the supergranules disappear near the centre of the solar disk (see Figure \ref{fig:supergranulation} for an example). Despite the slower speeds involved, the large spatial scale of supergranulation compared to granulation leads to larger disk-averaged RV variations. Various studies have estimated the amplitude of supergranulation signals in solar RVs to be in the range 0.5--1\,m/s using simulations \citep{2015A&A...583A.118M, 2019A&A...625L...6M, 2020A&A...642A.157M} and Sun-as-a-star observations \citep{2023A&A...669A..39A, 2024MNRAS.527.7681L}. This makes supergranulation the next largest stellar variability signal after magnetic activity in the Sun, and a very significant limiting factor in the RV detection of Earth-twins  \citep{2019A&A...625L...6M, 2020A&A...642A.157M}. The characteristic lifetime of supergranulation has been widely debated. Timescales ranging from 0.5 to 2 days have been reported in the literature \citep[see][for a review]{2018LRSP...15....6R}.  

Convection at the solar surface is strongly coupled to the Sun's magnetic dynamics. For example, \citet{1962ApJ...135..474L} and \citet{1964ApJ...140.1120S} found strong correlations between the magnetic field distribution of the quiet-Sun and supergranulation flows.  \citet{1988ApJ...327..964S} showed that supergranules are herded by small-scale magnetic structures at their boundaries. 
Given this link between magnetic field and supergranules, it is natural to ask how the physical properties of the latter vary over the Sun's magnetic cycle.

\begin{table*}
 	\centering
 	\caption{List of variations of supergranule sizes and timescale with magnetic field from the literature. Updated from \citet{2007A&A...466.1123M}. The first 6 references concern studies at a given time (spatial variations), while the last 6 reference studies cover the solar cycle or part of it. FT means Fourier Transform. A plus sign (+) means an increase in cell size or timescale with increasing activity levels, while a minus sign (-) means the opposite. An equal sign (=) means no variation was found.}
 	\label{tab: SG size}
 	\begin{tabular}{ c c c c c c } 
 		\hline 
        Reference & Data & Method &  Size &  Timescale & Comments\\
 		\hline
 		\citet{1970SoPh...13..292S} & Ca II images & autocorrelation & + &  & \\
        \citet{1988SoPh..117..343W} & magnetograms & autocorrelation & + & - & Large errorbars  \\
        \citet{1996SoPh..165..223W} & magnetograms & autocorreltation & + &  &   \\
        \citet{1997ApJ...481..988H} & Ca II K images & segmentation & = &  & \\
       \citet{2002SoPh..207...11R} & Ca II K images & autocorrelation & - & & \\
       {\citet{2007A&A...466.1123M}}  & magnetograms & segmentation & - & & \\ 
        \hline
        \citet{1981SoPh...71..161S} & Ca II K images & autocorrelation & - &  & Via latitude variations \\
       { \citet{1989A&A...213..431M}} & Ca II K images & 2D TF & + &  & \\
        \citet{1994SoPh..152..139K} & Ca II K images & segmentation & - &  & Via brightness intensity \\
       \citet{1995SoPh..158..213K} & magnetograms & autocorrelation & +  &  & FWHM of the autocorrelation curves \\
        {\citet{1999A&A...344..965B}} & Ca II K images & segmentation & - &  & Over 1 year only\\
       \citet{2004ApJ...616.1242D} & Doppler & segmentation & - & - &  2 time series\\
       \citet{2003ESASP.517...43G, 2004IAUS..223...41G} & Doppler & helioseismology & = & - & Week dependence\\
      {\citet{2008A&A...488.1109M}} & magnetograms & velocity field divergences & - & & $1\sigma$ detection \\
       \citet{2011ApJ...730L...3M}  & magnetograms \& Ca II K images & segmentation & + &  & \\
       \citet{ 2017ApJ...841...70C} & Ca II K images & watershed &  + &  & \\
       \citet{ 2017ApJ...844...24M} & Ca II K images &  watershed & + & & \\
       \citet{2022RAA....22d5006R} & Ca II K images & segmentation & - & & \\ 
        {\citet{2023AstBu..78..606S}}  & Ca II K images & segmentation &  & + & \\
       \hline
        & & & & & \\
 	\end{tabular}
 \end{table*}

Several studies have addressed this question using solar observations; we have attempted to summarize their results in Table~\ref{tab: SG size}, which is an updated version of a similar table presented in \citet{2007A&A...466.1123M}. The vast majority of studies focused on the sizes of supergranules, while a small number looked at their lifetimes.  All the studies were based on resolved images of the Sun. Despite the relatively similar types of data used, there are obvious disagreements between the results reported in the literature: some studies find supergranule sizes to be positively correlated with the Sun's activity level, others anti-correlated, and the same goes for their lifetimes. \citet{2007A&A...466.1123M} proposed a possible explanation for these conflicting results: they found that while the size of supergranules is anti-correlated with the magnetic field strength within the supergranulation cells, larger supergranules have a stronger network at their boundary, which can lead to a negative or positive correlation being reported depending on how the magnetic activity level is defined. This highlights the care needed to interpret the results summarised in Table~\ref{tab: SG size}.

While spatially resolved studies provide valuable insights into the physical processes governing supergranulation, its impact on RV planet searches is best studied using disk-integrated, Sun-as-a-star RV observations. \citet{2024MNRAS.527.7681L} recently used images from the  Solar and Heliospheric Observatory (SDO) to estimate the RV variations due to active regions, and subtracted them from HARPS-N RVs to study the residual 'quiet-Sun' RV variations. They used structure functions, a non-parametric method to quantify the root-mean-square (r.m.s.) variability of an irregularly sampled time-series as a function of timescale. They found that the r.m.s.\ of the quiet-Sun RVs is approximately constant over the 8 years of data they studied, at around 1m/s, and that its characteristic timescale is consistent with supergranulation simulations, indicating that supergranulation is the dominant phenomenon contributing to the residual variations. 

The present study aims to revisit this question and expand on this work, with a few important differences. As we have seen when discussing the published results based on spatially resolved images, the way magnetic activity is defined can have a critical effect on our understanding of the connection between supergranulation and magnetism. We therefore consider two different methods to estimate and subtract the RV contribution of the active regions, one based on SDO images (following \citealt{2024MNRAS.527.7681L}) and one using pixel-level correlations between the spectrum time-series and activity indicators \citep{Cretignier(2021)}. When using the latter, we consider an additional year of data, up to and including the present activity cycle maximum.  Finally, we model the supergranulation-like signal in the quiet-Sun RVs as a Gaussian Process (GP), following the method presented in \citet{2024MNRAS.531.4181O}, which allows us to measure its variance and characteristic timescale directly, with robust uncertainties. 

The remainder of this paper is structured as follows. Section~\ref{observations} presents the observations and data reduction steps used in this study as well as the two methods used to extract quiet-Sun RVs, while Section \ref{modelling} summarises the GP regression method we use to measure the (super)granulation parameters. We present our results and compare them with the relative sunspot number in Section~\ref{sec:results}. We then investigate in Section~\ref{observing strategies} different observing strategies that could be used by EPRV surveys to characterise supergranulation signals in other stars than the Sun. Finally, we summarise our conclusions and discuss the implications of our results in Section \ref{conclusion}.

\section{Observations}
\label{observations}

This study used observations from the HARPS-N solar telescope and the Solar Dynamics Observatory (SDO). We combine these data sets using different reduction methods to create two 'quiet-Sun' RV data sets. The raw observations and the methods to extract the quiet-Sun data are described below. 

\subsection{Raw Data}

\subsubsection{HARPS-N spectra and RVs}
\label{HARPS-N Data}

The HARPS-N Solar telescope, situated on the Telescopio Nazionale Galileo (TNG) at the Roque de las Muchachos Observatory in La Palma (Spain), has been observing the Sun since 2015 at a 5 minute cadence for (on average) 6 hours a day, allowing for a near continuous data set \citep[][]{2012SPIE.8446E..1VC, 2015ApJ...814L..21D, 2016SPIE.9912E..6ZP, 2019MNRAS.487.1082C, 2021A&A...648A.103D}{}{}. The telescope collects disc-integrated light from the Sun through a 3" lens that feeds into an integrating sphere, which scrambles the angular information, converting images to a point source. This in turn is fed through an optical fibre into the HARPS-N spectrograph, leading to a sun-as-a-star spectroscopic data set.  
These observations have very high resolution ($R = 115\,000$), and cover a wide wavelength range, from 383 to 690\,nm. The Sun is observed using 5-minute exposures, resulting in a typical SNR of 400 per resolution element (at 550\,nm).
We use HARPS-N solar data taken during the period from July 2015 to November 2023. As the 5-minute cadence matches the characteristic frequency of the Sun's $p$-mode oscillations, these are mostly averaged out, while the granulation with a timescale greater than five minutes, supergranulation and rotational modulation signals remain, along with instrumental systematics. 

The solar spectra collected were processed, and RVs extracted, using version 3.0.1 of the ESPRESSO Data-Reduction Software \citep[DRS;][]{2021A&A...645A..96P}{}{} optimized for HARPS-N \footnote{The ESPRESSO DRS is publicly available on \url{https://www.eso.org/sci/software/pipelines/espresso/?utm_source=chatgpt.com}.}, with a number of specific adjustments for the solar data as described in \citet{2021A&A...648A.103D}. The DRS data products used in the present work were the order-merged spectra (S1D) and the RVs extracted by cross-correlation with a digitised mask. 

\begin{figure*}
    \centering
    \includegraphics[width=0.98\textwidth]{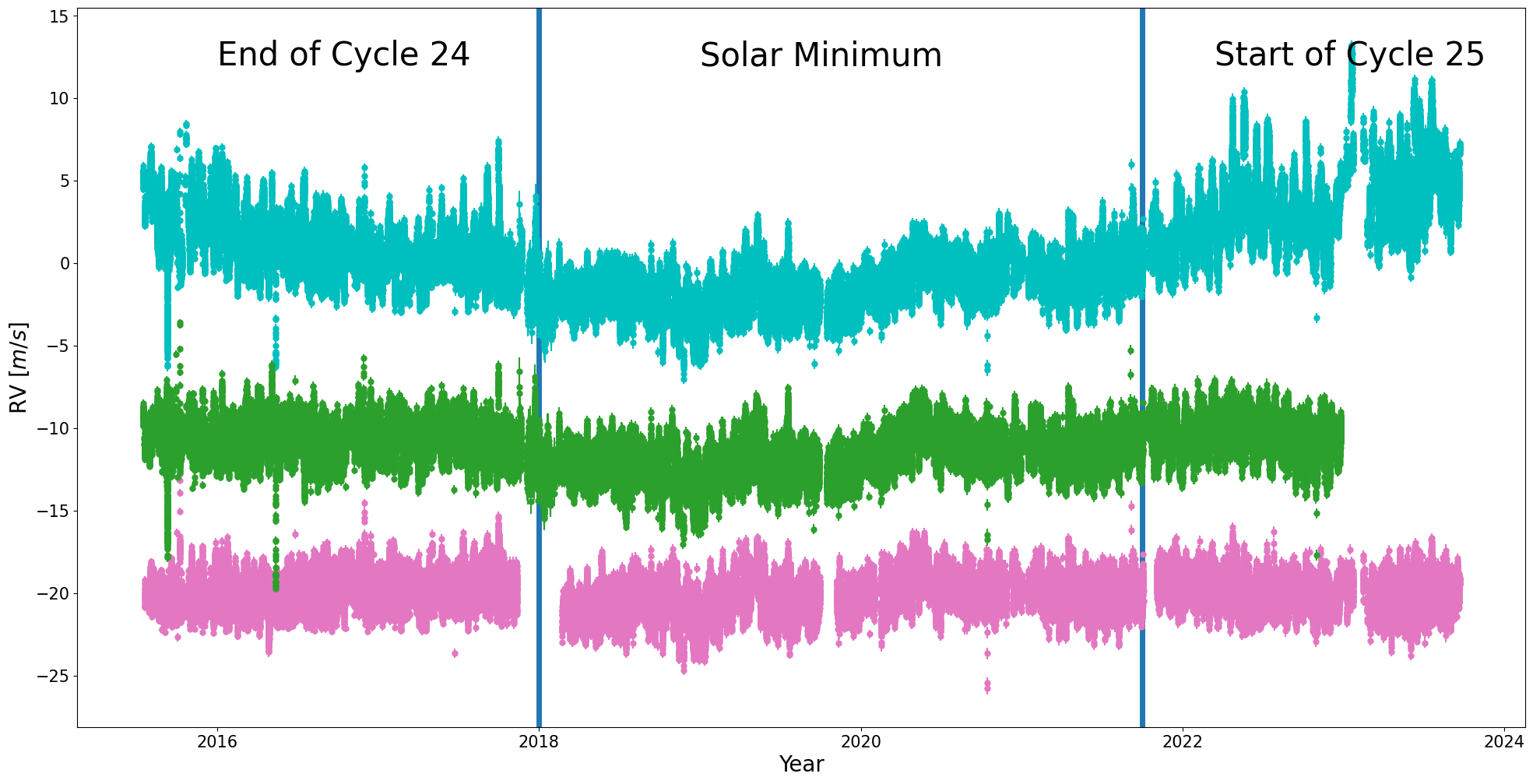}
    \caption{HARPS-N solar RVs (light blue, top), SDO quiet-Sun RVs (green, middle), and YARARA quiet-Sun RVs (pink, bottom), with arbitrary offset for graphical consideration. The vertical lines delimitate the end of cycle 24, the solar minimum, and the start of cycle 25. }
    \label{fig:all data}
\end{figure*}

\begin{figure*}
    \centering
    \includegraphics[width=0.98\textwidth]{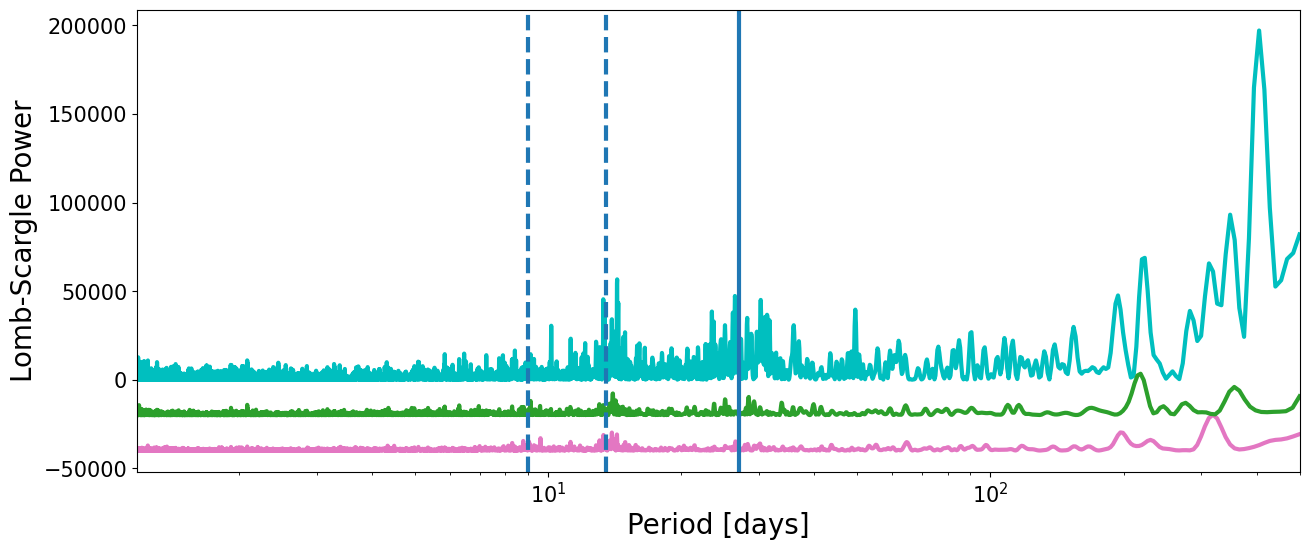}
    \caption{ Lomb-Scargle Periodogram of the HARPS-N Solar RVs (light blue), SDO quiet-Sun RVs (green), and YARARA quiet-Sun RVs (pink). The vertical lines indicate the solar rotation period and the first two harmonics (dashed).  }
    \label{fig:all data psd}
\end{figure*}

Observations affected by clouds or strong differential extinction across the solar disk were excluded following the prescription described in \citet{Klein(2024)}, which is based on the quality flags and differential extinction corrections derived by \citet[][]{2019MNRAS.487.1082C}{}{}. This selection resulted in a total of $107\,602$ solar spectra used in our analysis. On average, there are around 50 5-minute observations a day, over a 5.3 hour long observing day.  We show that this sampling does not affect our results in Appendix \ref{app: reduced}.  The HARPS-N solar RVs are shown in the top of Figure \ref{fig:all data}. The corresponding General Lomb-Scargle Periodogram \citep{2009A&A...496..577Z} is shown at the top of Figure \ref{fig:all data psd}.

\subsubsection{SDO/HMI Dopplergrams and magnetograms}
\label{SDO Data}

The Helioseismic and Magnetic Imager (HMI) onboard SDO has been collecting resolved images of the Sun since 2010, enabling detailed monitoring of the Sun's surface \citep{2012SoPh..275....3P}. 
HMI takes observations in two polarisation states, in each of 6 narrow bands around the magnetically sensitive 6176\,\AA  Fe {\sc I} line. 
A Gaussian function is fit to the intensities to estimate the local RV at each pixel and construct Dopplergrams, while the polarization information is used to determine the magnetic flux at each pixel and construct magnetograms. 

In the present work, we used $12\,343$ SDO/HMI observations, each with a 12-minute exposure, taken every 4 hours between July 29th 2015 to November 12th 2022. 

\subsection{Extracting quiet-Sun RVs}
\label{Quiet Sun RVs}

This study focuses on the RV variations of the quiet photosphere. To isolate these, we first need to estimate the contribution of magnetically active regions, which otherwise dominate the disk-integrated RV variations, and subtract this contribution from the HARPS-N solar RVs. 

We used two different methods to do this, one using SDO data and the other using HARPS-N data alone, allowing us to compare the results of the two approaches in the remainder of the paper.

\subsubsection{SDO activity correction}
\label{SDO actity}

We used SDO/HMI Dopplergrams and magnetograms to estimate and subtract the contributions of magnetically active regions to the HARPS-N disk-integrated solar RVs following the methodology developed by \citet{2016MNRAS.457.3637H, 2019ApJ...874..107M,2022ApJ...935....6H}. This is the same method used by \citet{2024MNRAS.527.7681L} to study the evolution of the quiet-Sun RV variations over multi-year timescales, allowing us to perform a direct comparison to the results of that study. 

This method involves first identifying which pixels on each HMI epoch belong to active regions by applying a magnetic flux   threshold in the magnetogram, and a minimum area requirement (i.e.\ only active regions above a certain size are considered). We started off using the thresholds defined in 
\citet{2016MNRAS.457.3637H} and \citet{2019ApJ...874..107M}, which were also used by \citet{2024MNRAS.527.7681L}. 
The magnetic threshold is defined by \citet{2016MNRAS.457.3637H} as 
\begin{equation}
    |B_{r, \mathrm{thresh}, ij}| = 24 G/\mu_{ij}
    \label{magnetic thresehold}
\end{equation}
where $\mu_{ij}$ is given by
\begin{equation}
    \mu_{ij} =\cos\theta_{ij}
    \label{magnetic thresehold nu }
\end{equation}
where $\theta_{ij}$ is the angle between the line of sight and the outward normal to the feature on the solar surface and the area threshold used was 20 micro-hemispheres. 

Once the active regions have been identified, the HMI Dopplergrams are used to estimate their contribution to the overall disk-integrated RVs. Owing to the presence of strong daily systematic effects in the HMI Dopplergrams, and the very different wavelength range covered by HMI and HARPS-N, this cannot be done by simply summing over the relevant pixels in the Dopplergrams. Instead, the estimation is done using a physically motivated, 2-component model that is empirically calibrated on the HARPS-N RVs. The first component of the model represents the photometric signature of active regions, which results from dark sunspots and bright plage rotating in and out of view as the Sun spins, breaking the symmetry in the disk's rotational profile. The second component corresponds to the localised suppression of the convective blueshift in regions of enhanced magnetic flux (this is the dominant component in the Sun). This procedure was used in the present work without modification, and we refer the interested reader to the relevant papers (\citealt{2016MNRAS.457.3637H, 2019ApJ...874..107M,2022ApJ...935....6H}) for a more detailed description.

Following \citet{2024MNRAS.527.7681L}, we subtract the SDO-estimated active-Sun RVs from the HARPS-N solar RVs to produce the quiet-Sun SDO RVs used in this study. Linear interpolation was used to bridge the cadence gap between the SDO and HARPS-N observations (4 hours \emph{versus} 5 minutes). \citet{2024MNRAS.527.7681L} showed that this does not affect the quiet-Sun results. The SDO quiet-Sun RVs and the corresponding periodogram are shown in Figures \ref{fig:all data} and \ref{fig:all data psd}.

\subsubsection{YARARA activity correction}
\label{YARARA activity}

Our second method uses YARARA \citep{Cretignier(2021)}, a post-processing methodology designed primarily to deliver improved RV precision compared to the DRS data products by correcting some instrumental systematics, as well as the effects of variable telluric absorption and stellar activity at the spectral level. Unlike the SDO activity correction, which makes use of resolved solar images, YARARA is fully data-driven, and relies mainly on the wavelength dependence of the effects being corrected (folding in prior knowledge such as the relevant reference frame.)

The starting point for YARARA post-processing is the time-series of 1-D, order-merged spectra produced by the DRS \citep{2021A&A...648A.103D}, re-interpolated onto a common wavelength grid. Spectra are daily stacked to increase the SNR of the observations; this is required for a reliable detection and correction of the various effects at the spectrum level. Furthermore, we would not be able to process more than $\sim 2000$ observations due to memory limitations (bearing in mind that each spectrum consists of $\sim 10^5$ pixels). First, the continuum is modelled and the spectra are continuum-normalised using RASSINE \citep{Cretignier(2020b)}. We then applied corrections for cosmic rays, tellurics, an interference pattern specific to HARPS-N, and ghosts (see \citealt{Cretignier(2021)} for details of each step). The correction of the instrumental Point-Spread Function (PSF) variations, first introduced in \citet{Stalport(2023)} and already applied to solar observations in \citet{Klein(2024)}, was also included to mitigate subtle effects such as the change in PSF following the replacement of the HARPS-N cryostat change, and variations in the projected rotational velocity ($v \sin i$) of the Sun as seen from the Earth \citep[first mentioned in][]{2019MNRAS.487.1082C}.
Finally, the YARARA activity correction is carried out by linear fitting and subtracting the dependence of the flux at each wavelength on the  Ca {\sc ii} H\&K $S$-index, as described in \citet{Cretignier(2021)}. Since the $S$-index variations are dominated by the contributions of plage/faculae \citep{Cretignier(2024)}, these are the main types of active regions we expect the YARARA activity correction to account for and contribution from spots and network are therefore still likely at the spectrum level. 
On the other hand, unlike the SDO correction, no explicit magnetic flux or area threshold is used in this correction.
After all these corrections have been applied to the spectra, the RVs are extracted by using the same cross-correlation procedure, with the same G2 mask, as was used in the ESPRESSO DRS. Note that this mask was produced by using a line list tailored to the Sun, where the line centers of the CCF mask were obtained from a parabola fit on the core of the lines as in \citep[][priv. communication]{Cretignier(2020a)}. CCFs are computed on the colour corrected spectra following the guidelines of \citet{Barragan(2024)} in order to remove any airmass dependency or change in atmospheric conditions \citep{Lovis(2007),Cretignier(2022b)}.  

Because YARARA is working with daily-binned data, the methodology is blind to effects or systematics that act on a timescale shorter than a day. In order to recreate the native 5-min sampling, we daily centered all the DRS RVs by the daily weighted mean before adding back the YARARA corrected daily binned RVs. In practice, such transformation is similar to a step-wise interpolation of the signals from the daily binned to the 5-min cadence. We slightly improved this extrapolation by assuming that the stellar activity signal should be smooth over a few days baseline. To do so, we interpolated linearly the YARARA activity model for the RVs and compute the difference with the step-wise interpolation. This extra correction vector was applied on the RVs. We refer to the resulting RVs as the YV1 RVs. 

The YARARA quiet-Sun RVs and their corresponding periodogram are shown in Figures \ref{fig:all data} and \ref{fig:all data psd}. The periodograms of the SDO and YARARA quiet-Sun RVs do not show significant peaks at stellar rotation period, indicating that the stellar activity has been effectively removed. We note that there is evidence of  activity residuals at around 14 days period (Prot/2) and, to a lesser extent, at around 8 days (Prot/3) in the activity-corrected RVs, weather the activity-correction was done with YARARA or SDO. In Appendix \ref{app: resids} we show that these residuals do not have an effect on the supergranulation signal that we model.


\section{Modelling the Quiet-Sun RVs with Gaussian Processes}
\label{modelling}

We follow the method described in \citet{2024MNRAS.531.4181O} to model the quiet-Sun RVs to measure the standard deviation and characteristic timescale of the remaining variability signals, and how these change over the Sun's activity cycle. We give a brief description of the method below; interested readers are referred to \citet{2024MNRAS.531.4181O} for a more detailed explanation. 

Stellar RV variations are typically modelled as the sum of several stochastic processes, some of which are quasi-periodic (such as $p$-mode oscillations and rotational modulation of active regions), while others are aperiodic (such as granulation and supergranulation). In this work, we include only the latter, as the 5-min exposure times of the HARPS-N solar data mostly average out the $p$-modes, while the activity signals have been suppressed in the manner described in Section~\ref{Quiet Sun RVs}. 

We use a two-component Gaussian Process (GP) model, where each component is an over-damped harmonic oscillator implemented in the {\tt celerite2} package \citep{celerite1,celerite2} as a Simple Harmonic Oscillator (SHO) kernel with a quality factor $Q=1/\sqrt{2}$. This kernel has a Power Spectral Density (PSD) given by:
\begin{equation}
    \label{eq:PSD_SHO_Qs2}
    P_{1/\sqrt{2}}(\nu) = \frac{2 \, S_0}{1+(\nu/\nu_0)^4},
\end{equation}
where $\nu_0$ is the undamped frequency of the oscillator and $S_0$ controls its amplitude. The variance of the resulting process is 
\begin{equation}
    \label{eq:variance}
    \sigma^2 = S_0 \omega_0 Q = \frac{1}{\sqrt{2}} S_0 \omega_0, 
\end{equation}
where $\omega_0$ is the undamped angular frequency given by $2\pi\nu_0$. It is also convenient to define the undamped period
\begin{equation}
    \label{eq: tau}
    \tau = 2\pi/\omega_0. 
\end{equation}
Here, $\tau$ corresponds to the timescale at which the PSD drops. This is the same definition as used by \citet{2023A&A...669A..39A}, however, in the {\tt celerite2} package this is defined as $\rho$.

The PSD in Eq ~\ref{eq:PSD_SHO_Qs2} is similar to the "Harvey law" \citep{1985ESASP.235..199H} that is frequently used to model solar and stellar granulation and supergranulation signals (see e.g.\  \citet{2023A&A...669A..39A}), but asymptotes to a steeper power-law (index 4 rather than 2) at high frequencies.  However, we note that there is ongoing debate in the literature as to the most appropriate power-law slope to use for convection-related signals (see e.g. \citealt{1997A&A...328..229N}).  


Our model also includes a white noise term added to the diagonal of the covariance matrix to account for the uncertainty due to photon noise and instrumental systematics.
Overall, the GP kernel we use to model the quiet-Sun RVs is then

\begin{equation}
    \label{eq:quiet RV Kernel}
    k_{\rm quiet} (\delta_t) = k_{\rm gran} (\delta_t) + k_{\rm SG} (\delta_t) + k_{\rm n}(t_i,t_j),
\end{equation} 

where $k_{\rm gran}$ and $k_{\rm SG}$ are aperiodic SHO kernels:
\begin{equation}
    \label{eq:k_SHO}
    k_{{\rm SHO},Q=1/\sqrt{2}} (\delta_t) = S_0 \, \omega_0 \, e^{-\frac{1}{\sqrt{2}} \omega_0 \delta_t} \, \cos \left( \frac{\omega_0 \delta_t}{\sqrt{2}} - \frac{\pi}{4} \right),
\end{equation}
with $\delta_t \equiv |t_i-t_j|$, representing the granulation and supergranulation components respectively, and $k_{\rm n}$ is a white noise term:
\begin{equation}
    \label{eq:k_w}
    k_{\rm n} (t_i,t_j) = \delta_{ij} (\sigma_{\rm n}^2 + \sigma_{i}^2),
\end{equation}
where $\sigma_w$ is the white noise standard deviation and $\sigma_{i}$ is the formal uncertainty of measurement $i$ .

A key goal of this study is to investigate how the granulation and supergranulation characteristics change over the Sun's magnetic cycle. To this end, we split the quiet-Sun RVs into 4 week chunks and model each chunk independently. This also minimizes the impact of any residual activity signals in the quiet-Sun RVs on our results, as the duration of the chunks is similar to the Sun's rotation period. Any 4 week chunks containing fewer than 100 observations (indicative of significant data gaps) were excluded from the analysis. However, on average the chunks had around 1000 data points. 

Following \citet{2024MNRAS.531.4181O}, we use a Markov Chain Monte Carlo (MCMC) to sample the joint posterior probability distribution over the 5 free parameters of our model, which are the $S_0$ and $\omega_0$ values for the granulation and supergranulation kernels, and the white noise standard deviation $\sigma_{\rm w}$. We used version 3 of the {\sc emcee} package \citep{2013PASP..125..306F,2019JOSS....4.1864F} to perform the MCMC sampling. The walkers are initialised in a tight Gaussian ball (with standard deviation 0.01\,dex) around a local optimum found using the {\sc minimize} function in {\sc scipy}'s {\sc optimize} module. Ln-uniform priors are used for all the parameters, within the interval $[-10;10]$,  in {$days^{-1}$ for $\omega$ and $(m s^{-1})^2$ days for $S$. The standard deviation of the white noise $\ln \sigma_{\rm n}$ is restricted to the interval $[-4;4]$ $m s^{-1}$. The number of walkers is set to 20 (4 times the number of parameters), and the MCMC chains are run for 50\,000 steps. To differentiate between the granulation and supergranulation hyper-parameters, we set the first  $\omega_0$ to be larger than the second, corresponding to the granulation signal. 

To assess the convergence of the chains and to choose the appropriate burn-in and thinning factors we estimate the auto-correlation length of the chains using {\sc emcee}'s built-in functionality. We consider the auto-correlation time estimates to be reliable as long as the longest auto-correlation length estimate across all parameters is $\tau_{\rm max} \le 200$ steps, i.e. \ less than a $50^{\rm th}$ of the chain. In this case we say the chains are well-converged. We discard the first $3 \tau_{\rm max}$ samples as part of the burn-in phase, and thin the remainder of the chains by a factor $\tau_{\rm max}/4$. In the cases where the chains do not converge in 50\,000 steps (due to large gaps in the data) we exclude the fit from future analysis. 

\section{Results}
\label{sec:results}

\begin{figure*}
    \centering
    \includegraphics[width=0.98\textwidth]{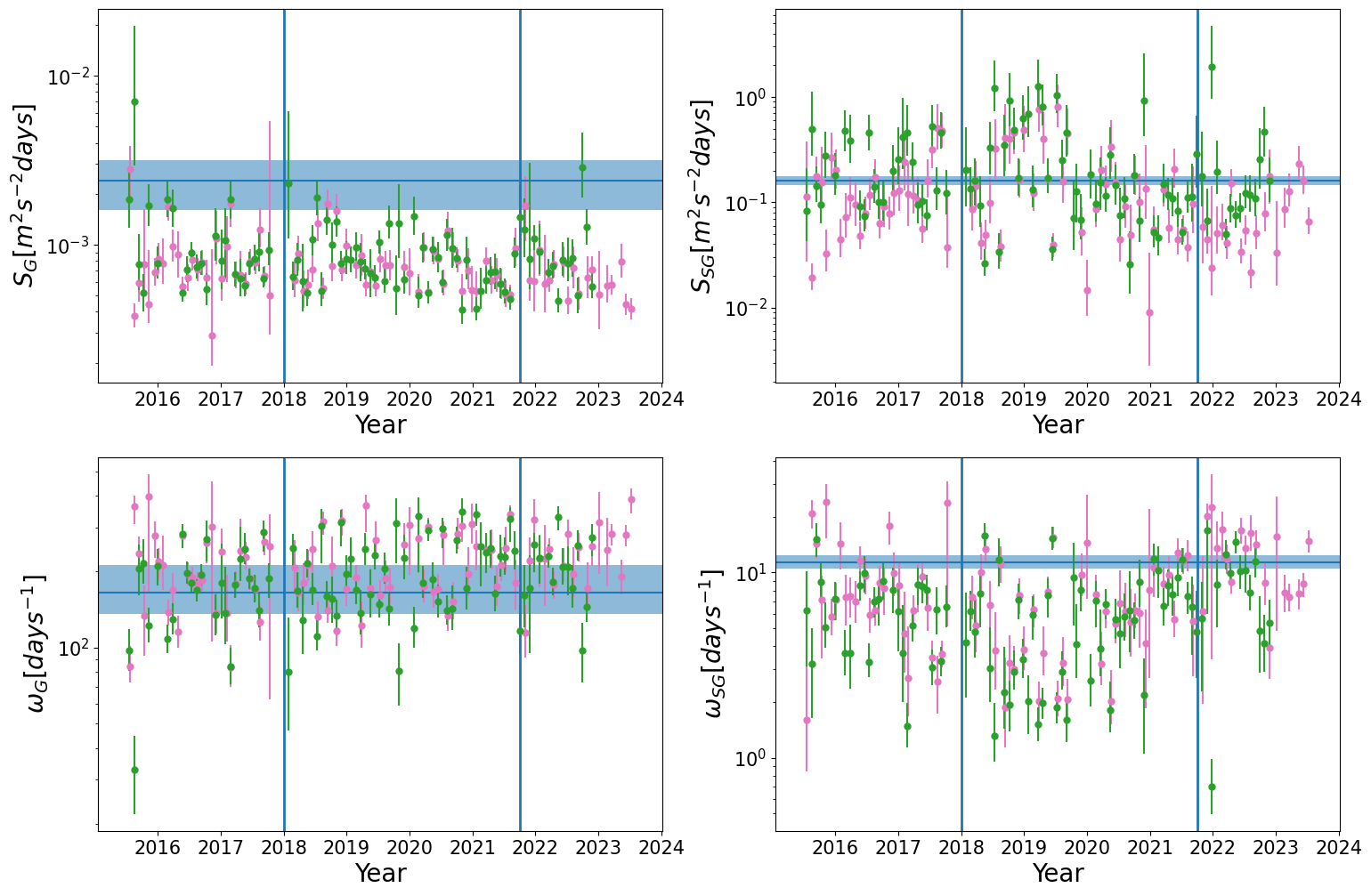}
    \caption{Parameters of the GP fits to the SDO and YARARA quiet-Sun RVs obtained in 4-week chunks over the solar cycle. The parameters of the granulation component, $\log S_{\rm G}$ and $\log \omega_{\rm G}$, are shown in the left hand column, while the supergranulation parameters  $\log S_{\rm SG}$ and $\log \omega_{\rm SG}$ are shown in the right hand column. The SDO and YARARA results are shown in green and pink, respectively. The blue horizontal line and shaded area represent the values and uncertainties reported for these parameters by \citet{2023A&A...669A..39A}. The blue vertical lines show the end of solar Cycle 24, the solar Minimum and the start of solar Cycle 25}
    \label{fig:YARARA + SDO  Results w+s}
\end{figure*}

The results of our fits for both the SDO and the YARARA quiet-Sun RVs are shown in Figure \ref{fig:YARARA + SDO  Results w+s}, where we also compare them with the published values for these parameters reported by \citet{2023A&A...669A..39A}, who analysed HARPS and HARPS-N solar data between 2015 - 2018. For ease of interpretation, we converted the $S_0$ and $\omega_0$ values for each component to the standard deviation $\sigma$ and undamped period $\tau$ of the process, these results are shown in Figure \ref{fig:YARARA + SDO  Results}.  Corner plots corresponding to the various stages of the solar activity cycle for both the SDO and YARARA results are given in Appendix \ref{app: corner plots }. 

\begin{figure*}
    \centering
    \includegraphics[width=0.98\textwidth]{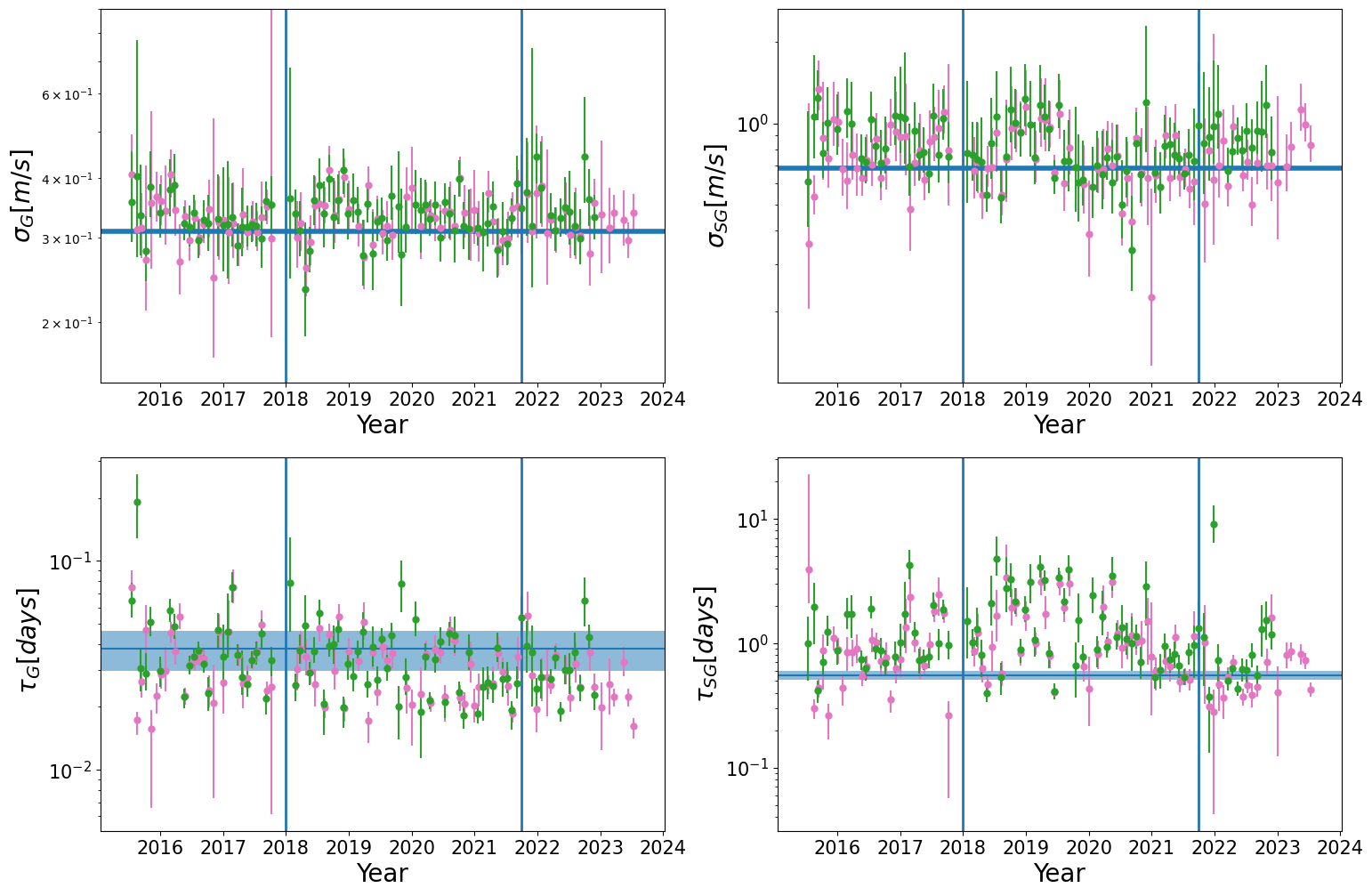}
    \caption{Same as Figure~\ref{fig:YARARA + SDO  Results w+s}, but converted to standard deviation $\sigma$ and time-scale $\tau$. The blue horizontal line and shaded area represent the values and uncertainties reported for these parameters by \citet{2023A&A...669A..39A}.}
    \label{fig:YARARA + SDO  Results}
\end{figure*}

Three outliers are apparent in both the YARARA and SDO results, with parameter values that differ significantly from the general trend and large uncertainties. These correspond to 4-week chunks that contained at least one large (several day) data gap, despite containing $>100$ points. These were discarded from the remainder of the analysis.

In Table \ref{tab: Average Values} we summarise the average values for the whole solar cycle, the end of Cycle 24, the solar minimum, and the beginning of Cycle 25. For the average SDO $\tau_{SG}$ for the beginning of Cycle 25, we remove the anomalous result of a time scale of 8.91 days. 

 \begin{table*}
 	\centering
 	\caption{Mean values for  $\sigma_{G}$, $\tau_{G}$, $\sigma_{SG}$, and $\tau_{SG}$ for the whole solar cycle, the end of Cycle 24, the solar minimum, and the beginning of Cycle 25. The values found by \citet{2023A&A...669A..39A} are also given. We note that the YARARA $\tau_{SG}$ values are systematically shorter than the corresponding SDO counterparts. This maybe due to the YARARA post-processing pipelines that correct for a number of different instrumental and telluric effects.}
 	\label{tab: Average Values}
 	\begin{tabular}{c c c c c c} 
 		\hline 
         & & $\sigma_{G}$ $[m/s]$ & $\tau_{G}$ $[days]$ & $\sigma_{SG}$ $[m/s]$  &$\tau_{SG}$ $[days]$\\
         \hline
    
         \multirow{2}{5em}{Whole  Cycle} & YARARA & $0.329 \pm 0.003$ & $0.031 \pm 0.001$ & $0.752 \pm 0.020$ &   $1.053 \pm 0.080$ \\
         & SDO & $0.336\pm0.004$ & $0.037 \pm 0.002$ & $0.834\pm0.020$ & $1.458 \pm 0.136$\\
 		\hline
        \multirow{2}{5em}{End of Cycle 24} & YARARA & $0.324 \pm 0.006 $ & $0.034 \pm 0.003$ & $0.794\pm0.036$ & $1.020\pm0.139$\\
         & SDO & $0.333 \pm 0.006$ & $ 0.044 \pm 0.006$ & $0.893\pm0.033$ & $1.245 \pm 0.149$ \\
         \hline
        \multirow{2}{5em}{Solar Minimum} & YARARA & $0.333 \pm 0.005$ & $0.031 \pm 0.001$ & $0.727\pm0.030$ & $1.274\pm0.125$\\
         & SDO & $0.332\pm 0.005$ & $0.035 \pm 0.002$ & $0.783\pm0.029$ & $1.597 \pm 0.163$\\
        \hline 
 	 \multirow{2}{5em}{Beginning of Cycle 25} & YARARA & $0.328\pm 0.006$ & $0.028\pm0.002$ & $0.740 \pm 0.037$ & $0.614 \pm 0.071$\\
         & SDO & $0.352\pm0.012$ & $0.033 \pm 0.003$ & $0.892\pm 0.034$ & $0.796 \pm 0.097$ \\
        \hline 
         \multirow{2}{5em}{\citet{2023A&A...669A..39A} }&  & \multirow{2}*{$0.31$} & \multirow{2}*{$0.038\pm0.008$}  &  \multirow{2}*{$0.68$} & \multirow{2}*{$0.554\pm0.046$}

        \\
        \\
        \hline
 	\end{tabular}
 \end{table*}

\subsection{4-week chunk granulation and supergranulation results}

The granulation parameters appear consistent over the Sun's activity cycle, as expected, see, for example, \citet{2020A&A...636A..70S}. The values measured from the SDO and YARARA-corrected RVs are approximately consistent with each other. While the timescale is consistent with the literature values, the amplitude $S_0$ is somewhat smaller. We note that the timescale found for granulation, on average 44 minutes for the YARARA results, is longer than the lifetime of the granules on the stellar surface. This is because we are looking at the characteristic timescale, corresponding with the knee of the granulation PSD, which in turn is attributed to the broad dispersion of granulation turnover timescales \citep{2011A&A...532A.108S}. 44 minutes is consistent with the literature values found using Harvey models \citep[e.g.][] {2023A&A...669A..39A}.  

On the other hand, the supergranulation parameters display significant variations over the solar cycle in both the YARARA and SDO results. This is promising, as both methods have limitations that may affect the supergranulation parameters. The SDO results may still include small active regions which fall below the area threshold of 20 micro-hemispheres defined in \citet{2019ApJ...874..107M}. On the other hand, the YARARA activity correction might remove not only activity, but also some of the supergranulation signals. We recall that the YARARA activity correction consists in a simple linear decorrelation between the measured flux at each wavelength (in the stellar rest-frame) and the Mount Wilson chromospheric activity index $S_{\rm HK}$. If the supergranulation signal is correlated with the chromospheric index, some of it may be removed in the process. As the two methods are in agreement, and their limitations work in opposite directions, we therefore conclude the effects of the limitations seem minimal, and the cycle we see is the result of a physics effect. 

To test for differences in the distributions of the parameters of the granulation and supergranulation models, we employ the use of Kolmogorov-Smirnov tests. We find that KS tests suggest that the supergranulation timescales found during the End of cycle 24, the minimum, and the beginning of cycle 25 are drawn from different distributions compared to each other, with p-values of $10^{-11}$ or less in both the YARARA and SDO case. We also find that when comparing the parameters found via the two different methods directly that $\sigma_G$, $\tau_G$, and  $\tau_{SG}$ are likely drawn from the same distribution. On the other hand, the KS test suggests that $\sigma_{SG}$ comes from a different distribution in the YARARA and SDO case (p-value  = $0.0002$). We propose that this difference comes from the different systematic attenuation and the differences in activity attenuation between the two methods.

\subsection{Short-term scatter in the 4-week chunk results}

\begin{figure}
    \centering
    \includegraphics[width=0.48\textwidth]{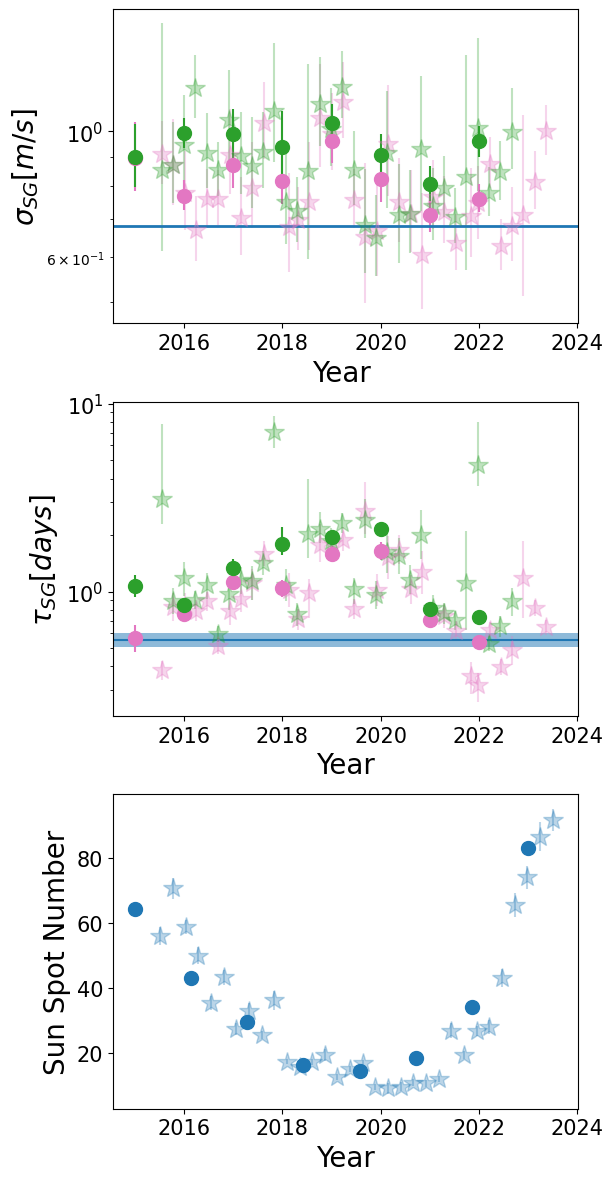}
    \caption{Supergranulation parameters obtained from SDO and YARARA quiet-Sun RVs when fitting longer chunks of 12 weeks (stars) and 1 year (circles). The colour scheme is the same as for Figures~\ref{fig:YARARA + SDO  Results w+s} and \ref{fig:YARARA + SDO  Results}, and the horizontal line and shaded area once again shows the literature values from \citet{2023A&A...669A..39A}. Also shown for comparison are the average Sunspot numbers over the same time frame.}
    \label{fig:Sunspot comp}
\end{figure}

Another important feature of our results, particularly for the supergranulation component, is the large amount of scatter in the measured parameter values shown in Figures~\ref{fig:YARARA + SDO  Results w+s} and \ref{fig:YARARA + SDO  Results}, especially for the SDO results. Leaving aside any long-term trends, the time-scales in particular vary by up to an order of magnitude between consecutive 4-week data chunks, and this scatter is significantly in excess of the formal uncertainties. To check that this was not a fortuitous effect for the specific set of 4 week long chunks used, we shifted all the chunk boundaries by 14 days and re ran the analysis, but the results were largely the same. We then repeated our analysis using longer 3-month (12-week) and 1-year chunks, and found that most of the scatter disappeared. The results are shown for the supergranulation component in Figure~\ref{fig:Sunspot comp}. Based on our experience using injection-recovery tests on simulated data \citep{2024MNRAS.531.4181O}, we believe that the scatter in the 4-week chunk results is real, and is not an artefact of the limited duration of the chunks. We therefore note that caution should be applied when interpreting supergranulation signals, even in observing campaigns lasting a month or more. The residuals in the SDO-corrected case have slightly larger scatter and contain more outliers than in the YARARA-corrected case. A possible explanation for this is that the YARARA post-processing corrects some HARPS-N instrumental systematics, which are not accounted for in the SDO-corrected data, in addition to the different ways in which the variations due to active regions are corrected.

\subsection{Variation of the supergranulation timescale over the solar cycle}

\begin{figure*}
    \centering
    \includegraphics[width=0.98\textwidth]{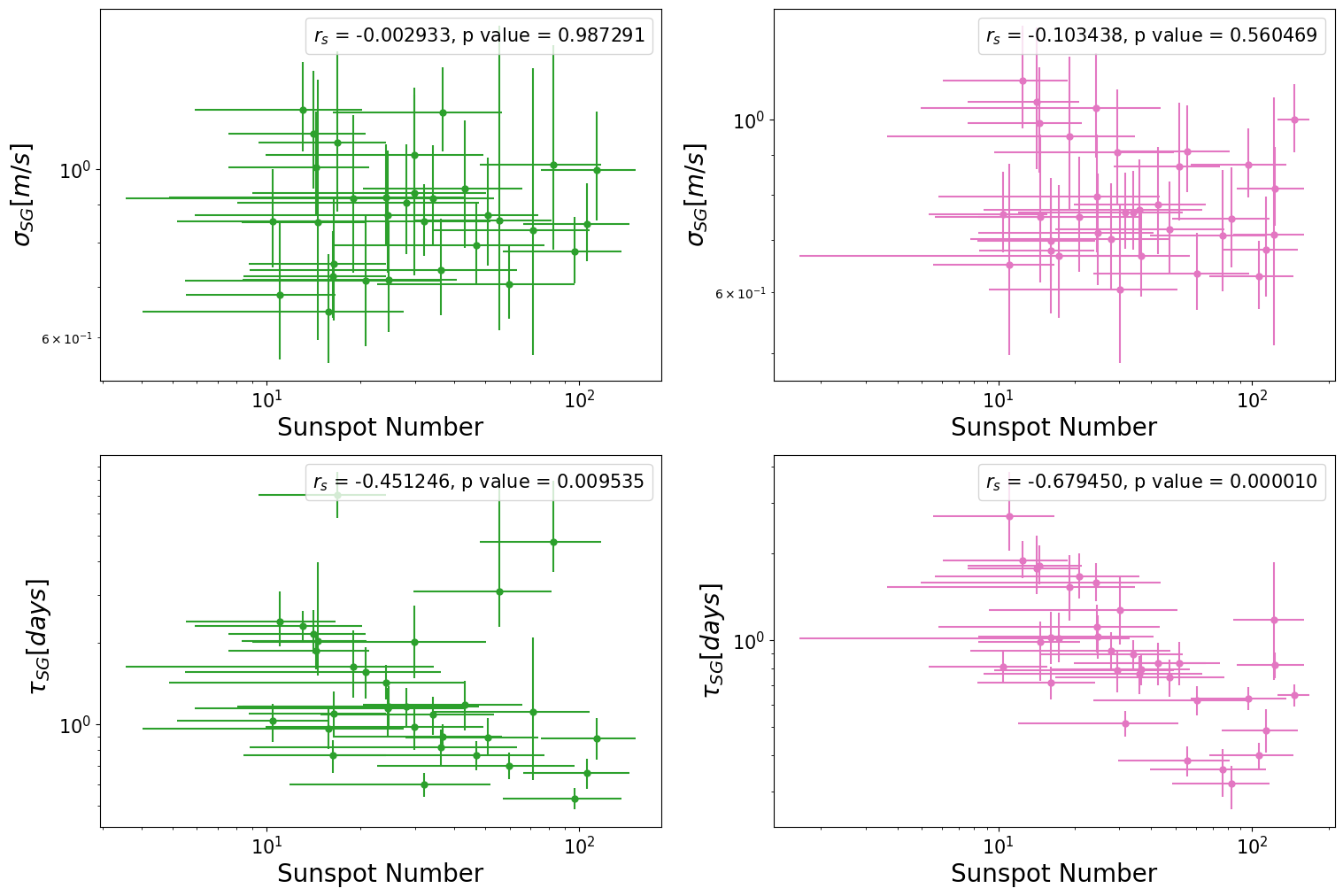}
    \caption{Correlation between the supergranulation parameters obtained from the 12-week (3-month) chunks and the Sun-spot numbers, for the SDO-corrected RVs (left, green) and the YARARA-corrected RVs (right, pink). In each panel we also report the $p$-values derived from a Spearman's rank correlation analysis, which correspond to the probability of a similarly correlated dataset arising from white Gaussian noise only.}
    \label{fig:p values 3}
\end{figure*}

The supergranulation timescale $\tau_{\rm SG}$  shows a clear variation over the decade spanned by our data. The variations are approximately in anti-phase with the average relative Sunspot number in the corresponding time-chunks, which we downloaded from the Royal Observatory of Belgium's Sunspot Index and Long-term solar Observations (SILSO) portal\footnote{See \url{https://www.sidc.be/SILSO/home.}} \citep{sidc}, and which is also shown in Figure~\ref{fig:Sunspot comp}. 

To quantify this correlation, we plotted in Figure~\ref{fig:p values 3} the supergranulation parameters measured from 3-month chunks as a function of average relative Sunspot number. We also computed the Spearman's rank correlation between each parameter and the Sunspot number, and report the corresponding $p$-values on Figure~\ref{fig:p values 3}. In Appendix \ref{p_value appen} we show the same for the 4 week and year long chunks.  We observe strong anti-correlations ($p$-value below $10^{-2}$) between the time-scale from the SDO-corrected RVs. The correlation is even stronger ($p$-value below $10^{-4}$) for the timescale derived from YARARA-corrected RVs. Both methods show very weak correlation for the standard deviation. We conclude that the time-scale of the supergranulation component displays a clear variation over the Sun's activity cycle, irrespective of the method used to evaluate it.



\section{Observing supergranulation in other stars}
\label{observing strategies}

Supergranulation has become one of the main issues in EPRV planet searches, but we do not yet know how it scales as a function of stellar parameters. Given that supergranulation is a convective process, we would expect it to decrease in magnitude towards later spectral types, and the limited sample of stars with asteroseismic measurements spanning 5-8 nights analysed by \citet{2011A&A...525A.140D} appears to confirm this. However,  magnetic activity level also plays a role, as shown by the present study in the case of the Sun.

A number of observational efforts have been proposed to attempt to characterise supergranulation in stars other than the Sun, but it is very challenging as the signal of interest has characteristic timescales similar to an Earth day. In this Section, we set out to evaluate the amount of observing time needed to adequately characterise this timescale and how it depends on the choice of observing strategy. 

Our goal is to estimate, for different observing strategies, the minimum duration of an observational campaign that would allow us to measure the  supergranulation parameters of each target star to a given precision. 

One possible way to do this would be to simulate large numbers of time-series with different parameters and time-sampling, then model them with a GP, sampling the posterior to evaluate the uncertainties, as we did for the solar data in Section~\ref{modelling}. However, this would be computational expensive. While we did this for a few simulations, we noted that it was hard to quantify the success of the simulation, and took significant time to test one strategy. We therefore adopt a Fisher information approach to the problem, following the work of \citet{2024AJ....168...29G}. For data drawn from a multivariate Gaussian distribution with mean vector $\boldsymbol{\mu}$ and covariance matrix $\mathbf{C}$, the elements of the Fisher information matrix $\mathbf{F}$ are given by:

\begin{equation}
\label{eq:fisher}
    \boldsymbol{F}_{i,j} = \left(\frac{\partial \boldsymbol{\mu}}{\partial\theta_i}\right)^T \mathbf{C}^{-1} \left(\frac{\partial \boldsymbol{\mu}}{\partial\theta_j}\right)+ \frac{1}{2} tr\left(\mathbf{C}^{-1} \frac{\partial \mathbf{C}}{\partial\theta_i} \mathbf{C}^{-1} \frac{\partial \mathbf{C}}{\partial\theta_j}\right).
\end{equation}

where $\boldsymbol{\theta}$ is a vector of parameters controlling the mean and covariance functions which produce $\boldsymbol{\mu}$ and $\mathbf{C}$. The parameter uncertainties are then given by:
\begin{equation}
    \sigma^2_{\theta_i} =\boldsymbol{F}_{i,i}.
\end{equation} 
Here we refer to the  Fisher information matrix as  $\mathbf{F}$, as opposed to $\mathbf{B}$, used by \citet{2024AJ....168...29G}, in order to avoid confusion with the magnetic threshold. 

In the context of RV surveys, the mean function encodes the planet signals, while the covariance function encodes the noise (photon, instrumental and stellar noise). The goal of \citet{2024AJ....168...29G} was to assess the impact of observing strategy on the characterisation of planets. They therefore worked with a fixed covariance matrix $\mathbf{C}$, and focused on the first term in Eq. ~\ref{eq:fisher}. In this work, we are interested in the impact of observing strategy on the characterisation of a stochastic signal. We therefore assume a trivial mean function, and the Fisher information matrix reduces to Eq. ~\ref{eq:fisher}:
\begin{equation}
\label{eq:fisher2}
    \boldsymbol{F}_{i,j} =  \frac{1}{2} tr\left(\mathbf{C}^{-1} \frac{\partial \mathbf{C}}{\partial\theta_i} \mathbf{C}^{-1} \frac{\partial \mathbf{C}}{\partial\theta_j}\right),
\end{equation}
where the elements of the covariance matrix are set to the sum of two terms given by Eq. \ref{eq:k_SHO}: one for granulation, and one for supergranulation (we implicitly assume that activity signals occur on a long timescale compared to the simulated observations, or are modelled separately). The derivatives of the covariance matrix are given by:

\begin{equation}
\label{eq:s_derivative }
    \frac{\partial C}{\partial S} = \omega e^{-\frac{\omega\tau}{\sqrt{2}}} \cos \left(\frac{\omega\tau}{\sqrt{2}} - \frac{\pi}{4}\right),
\end{equation}

\begin{equation}
\label{eq:w_derivative }
    \frac{\partial C}{\partial \omega}  =   \frac{- S  e^{-\frac{\omega\tau}{\sqrt{2}}}}{\sqrt{2}}\left( \omega \tau \sin \left( \frac{\omega\tau}{\sqrt{2}} -  \frac{\pi}{4}  \right)  + (\omega\tau -\sqrt{2})\cos \left(  \frac{\omega\tau}{\sqrt{2}} -  \frac{\pi}{4} \right) \right),
\end{equation}
We also include a 0.3\,m/s white noise term on the diagonal of the covariance matrix, representative of the expected precision for (e.g.) the HARPS3 instrument. One caveat of the Fisher Information approach is that it assumes that the real granulation and supergranulation signals are perfectly represented by the GP kernels. Of course, this may not be strictly true in practice, so the parameter uncertainties computed according to Eq.~\ref{eq:fisher2} should be treated as approximate. 

\begin{figure*}
    \centering
    \includegraphics[width=0.98\textwidth]{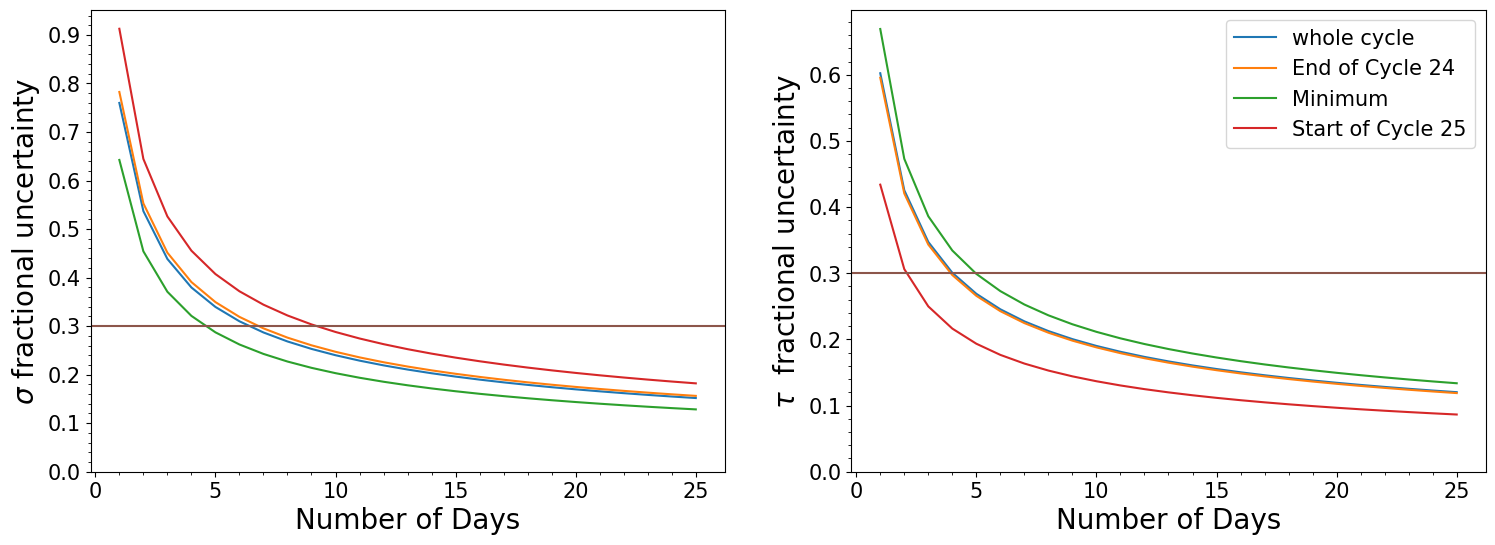}
    \caption{Expected $\sigma$ and $\tau$ fractional uncertainty as a function of number of nights of observation for continuous observations of a single star. The fractional uncertainty is calculated for various $\sigma$ and $\tau$ values corresponding to different phases of the solar cycle. The hyperparameters used are the average values shown in Table \ref{tab: Average Values}. The brown horizontal line corresponds to a fractional uncertainty of 0.3. }
    \label{fig:fisher cycle}
\end{figure*}

We started by considering the simplest possible observing strategy, where one star was monitored continuously every night for a number of nights. Each observation consisted of a 5-minute block. For simplicity we assume that every night lasts 8 hours, and ignore the impact of weather. Figure \ref{fig:fisher cycle} shows how the relative precision on the standard deviation and time-scale of supergranulation improves as the duration of the observing campaign increases, for different assumed values of the parameters (corresponding to the values measured for different parts of the solar cycle in Section~\ref{modelling}). Figure~\ref{fig:fisher cycle} demonstrates that at least a week or two of observations are needed to obtain any meaningful constraints on the supergranulation parameters. The time needed to reach a given precision depends on the time-scale of the signal, but only fairly weakly (a factor $\sim 2$ for the different curves shown, which span an order of magnitude in timescale). Finally, the relative precision curves fall exponentially, and the benefit of extending the observing campaign beyond one or two $e$-folding times becomes minimal. We note that while it takes the shortest amount of time to constrain $\sigma_{SG}$ at solar minimum, as we would expect, it takes the longest time to constrain  $\tau_{SG}$ during this period. This is due to the increase in $\tau_{SG}$ that we observe at solar minimum. 

To test that the Fisher information matrix uncertainties are accurate, we simulated and modelled 5 days of RV observations. The fractional uncertainty we obtained is similar to that calculated using the Fisher information matrix, indicating that the method is reliable. 

Next, we considered a more-realistic scenario where the observer cycles randomly between a fixed number of $N$ target stars, still using 5-minute exposures, but varying the number $M$ of consecutive exposures of each star taken before switching to the next star. This is intended to explore the trade-off between staying on a given star for longer, which provides better sampling of the granulation signal, versus spacing out the observations of each star in a given night, which improves the sampling of the supergranulation signal. Each time we switch to a new star, we include a pessimistic 5-min overhead for slewing and target acquisition. In actual fact, the overhead time is likely to be less. In this simulation, we also include an additional 0.1 m/s white noise to account for the oscillations. \citet{2019AJ....157..163C} showed that with a 5 minute exposure time, this would be the noise that oscillations would induce in a Sun-like star. For cooler stars, this would be lower.  Having simulated a set of time-stamps for each combination of $N$ and $M$, we then used the Fisher information matrix to estimate how many nights would be needed to measure the supergranulation timescale $\tau$ to a fractional uncertainty of $0.3$. Since the individual time-stamps are random, we repeated the process 100 times for each observing strategy. As we are trying to see how many stars we can characterise supergranulation in a month, we have an upper limit of 50 days to the test.

A $0.3$ fractional uncertainty limit was chosen arbitrarily as from Figure \ref{fig:fisher cycle} we see that reducing the fractional uncertainty beyond this limit would require significantly more survey time. Also, this is the level of detection we would need to be able to detect the supergranulation time scale variance over the solar cycle in the sun. While this level of uncertainty will not tell us the supergranulation time scales of other stars down to great precision, it will allow us to see if the time scales varies as we would expect it to, which would be a great improvement on our current understanding of supergranulation.


\begin{figure*}
    \centering
    \includegraphics[width=0.98\textwidth]{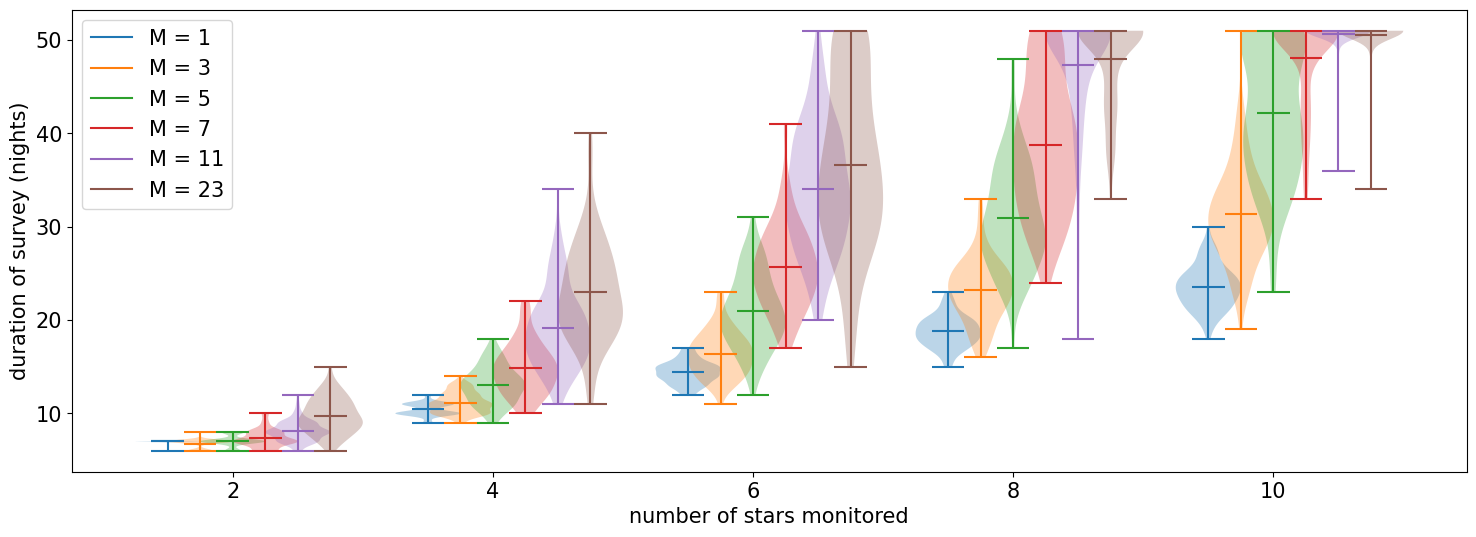}
    \caption{Number of nights needed to reach a fractional uncertainty of 0.3 in the supergranulation timescale $\tau$ as a function of the number of stars observed each night. Different colours correspond to different observational strategies, as described in the text. }
    \label{fig:fisher}
\end{figure*}


The results are shown in Figure \ref{fig:fisher}, where different colours represent different values of $M$. The first take-away point from that figure is that for a given number of stars per night, the time taken to reach the desired precision is systematically lower for $M=1$ (blue symbols on Fig~\ref{fig:fisher}), i.e.\ when the target star changes every 10 minutes, despite the fact that 50\% of each night is lost to overheads in this scenario. For example, this rapid switching approach would allow the characterisation of the supergranulation signal (albeit only at the modest precision of 30\%) to be completed in 23 nights for 10 stars. By contrast, staring at each star for an hour ($M=11$, shown in purple on Fig~\ref{fig:fisher}), this precision is only reached after 40 nights on a smaller target sample of $N=6$ stars.  The second take-away is that we can get roughly similar characterisation results by observing 2 groups of 4 stars for $\sim10$ nights each, compared to 8 stars observed across $\sim20$ nights.

    \label{fig:2 half nights}

\section{Discussion and conclusions}
\label{conclusion}

We used GP regression to measure the standard deviation and characteristic timescale of the granulation and supergranulation component of the HARPS-N solar RVs, and their evolution over the Sun's magnetic activity cycle, after removing the contribution of magnetically active regions using two different methods: one using resolved magnetograms and Dopplergrams from SDO, following \citet{2024MNRAS.527.7681L} and references therein, and one using the HARPS-N spectra themselves, following the YARARA methodology \citep{Cretignier(2021)}. 

Our results for granulation were consistent across the two datasets used, and also consistent with literature values after accounting for the 5-minute cadence of the observations. Furthermore, the granulation parameters appear stable over the full time-span of our dataset, which matches theoretical and observational expectations \citep{2018A&A...616A..87M, 2020A&A...636A..70S}. 

On the other hand, we have shown that the supergranulation properties, particularly its time-scale, vary by a factor of three over the Sun's activity cycle, with peak to peak variations up to a magnitude. Specifically, we observe a longer timescale at activity minimum compared to the rising and decaying phases of the cycle, and show that the supergranulation timescale is strongly negatively correlated with the average relative Sunspot number measured at a given time. 

Interestingly, \citet{2024MNRAS.527.7681L}, who studied the same SDO-corrected dataset using structure functions rather than GPs, did not observe significant changes in the structure functions over the Sun's activity cycle. This may arise from the daily sampling of the data, which makes the characterisation of signals on $\sim$ daily timescales, such as supergranulation, particularly challenging using a non-parametric method such as structure functions. By comparison, the GP regression approach used in the present work makes stronger assumptions about the nature of the signal being modelled. However, as discussed in \citet{2024MNRAS.531.4181O}, these assumptions are consistent with the common practice in the literature of using Harvey functions to model stochastic signals in stellar RVs.

To enable a direct comparison of our results to those of \citet{2024MNRAS.527.7681L},  we analytically computed the structure functions corresponding to the 1 year results for both the SDO and YARARA quiet-Sun RVs. 
For a  stationary time series exactly described by a known covariance function, $K(\tau)$, as is the case for Gaussian Processes, the corresponding structure function is given by
\begin{equation}
    \mathrm{SF}(\tau) = 2 \left (\sigma ^ 2 - K(\tau) \right). 
\end{equation}
These structure functions are shown in Figure \ref{fig:SF}.  
Following \citet{2024MNRAS.527.7681L}, we quantify variability as $\sqrt{\frac{1}{2}\mathrm{SF}}$.

At $\tau = 10^{-1}$ d, we see that there is a difference between the structure functions calculated at solar minimum and at more activity times, however by $\tau = 10^{0}$ d, this correlation is no longer present. The same applies to the structure functions found in \citet{2024MNRAS.527.7681L}, indicating that the results are consistent with each other.

\begin{figure*}
    \centering
    \includegraphics[width=0.98\textwidth]{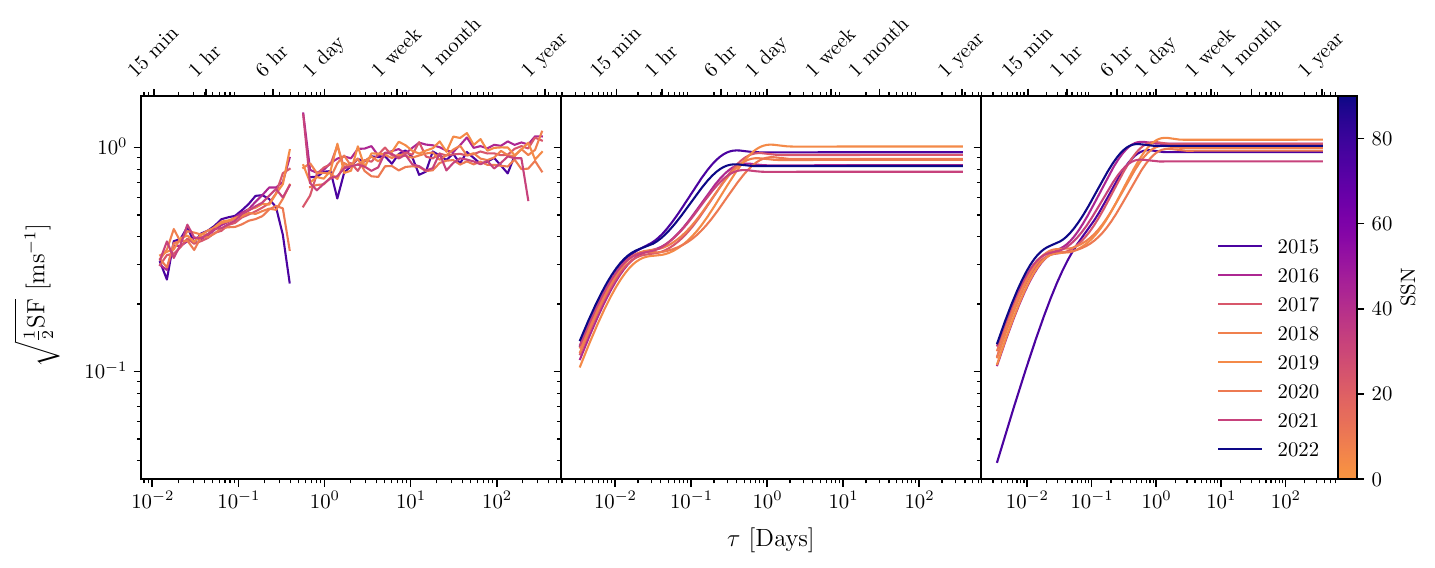}
    \caption{Structure functions corresponding to  the YARARA (middle) and SDO (right) hyperparameters calculated using year long chunks. The lines are coloured by the average sun spot number that year. The original results from \citet{2024MNRAS.527.7681L} are shown on the left.}
    \label{fig:SF}
\end{figure*}

Finally, we used a Fisher information approach to compute the number of nights needed to provide useful empirical constraints on supergranulation properties for stars other than the Sun. We showed that a dedicated survey spanning around 20 consecutive nights using an instrument with RV precision of order 30cm/s, such as HARPS3, can characterise the supergranulation timescale to a precision of around 30\% for up to 10 stars. This would lead to a significant improvement in our understanding of how supergranulation depends on stellar parameters such as temperature, surface gravity and metallicity, as well as activity level. Such a survey is currently being considered as part of the science verification phase of the Terra Hunting Experiment. 

On the other hand, we also saw that the supergranulation timescales measured for the Sun in 4-week chunks display considerable scatter over and above the formal uncertainties, which disappears when using 12-week chunks. The origin of this scatter needs to be investigated in more detail, as it may affect the interpretability of the results of such a "supergranulation" survey. 

As a sanity check we computed the residuals of the best-fit GP model in each chunk, divided them by the combined (model $+$ measurement) uncertainties, and checked that the results are consistent with a zero-mean, unit variance Gaussian distribution, which was always the case (irrespective of activity correction method or chunk duration). We also checked if the jitter term (white noise term added to the diagonal of the covariance matrix), which we fit alongside the supergranulation and granulation parameters, varies over the Sun's activity cycle, but it was constant within the uncertainties, with typical values in the range 0.20 -- 0.25\,m/s. The jitter term absorbs any variations not accounted for by the GP model, and this check shows that the GP model's ability to explain the data is quite good, and is not noticeably variable over the solar cycle. This increases our confidence in the fact that the change in the supergranulation timescale we observe over the solar cycle is not an artefact of the used model
(SHO kernel).

\subsection{What causes the "supergranulation cycle"?}

We have brought to light a clear dependence of the supergranulation time-scale on the magnetic activity state of the Sun. What could be its origin, and does it match theoretical expectations? 

As discussed in Section~\ref{sec:intro}, there is considerable divergence in the literature on how supergranulation changes as a function of magnetic activity in the Sun. Around half of the papers listed in Table~\ref{tab: SG size} find that the size of super-granulation cells is positively correlated with activity level, while the rest find a negative correlation. The literature on supergranulation cells' lifespan is much sparser, but  \citet{2004ApJ...616.1242D} and \citet{2023AstBu..78..606S} suggest that larger cells have a longer lifespan. Following this logic, the negative correlation we have evidenced between the supergranulation time-scale in the solar RVs and the Sunspot number would indicate that supergranulation cells tend to be larger at solar minima. In other words, our results would be consistent with the studies which have found a negative, rather than positive, correlation between cell size and magnetic activity level. It is also interesting to note that \citet{1981SoPh...71..161S} and \citet{2008A&A...488.1109M} suggest that the network magnetic elements have a limiting effect on supergranule sizes. If this is the case, we would expect larger cells at solar minimum, and thus longer timescales, which is consistent with our results. However, the overall changes in supergranule size over the Sun's activity cycle found by e.g.\ \citet{2007A&A...466.1123M,2008A&A...488.1109M} are too small (around 10\%) to explain the factor of two change in timescale we report in the present work. The cause of that timescale change therefore remains unexplained for the time being.



\subsection{Impact for RV surveys}

The RV signature of supergranulation has a significant impact on the sensitivity of RV surveys to small, temperate planets \citep{2020A&A...642A.157M}. It introduces correlations between observations taken in a given night which, if not modelled explicitly, constitute a fundamental night-to-night noise floor. If the supergranulation timescale is well understood, explicitly modelling the covariance of this signal could enhance the sensitivity to planet signals. We intend to quantify this using injection-recovery tests in a forthcoming paper. 

However, an important consequence of the results presented in this work is that the RV supergranulation signal cannot be treated as stationary if the observational baselines are on the order of an activity cycle. It is therefore important to a) continue investigating the correlation between supergranulation properties and activity level, b) develop strategies to characterise supergranulation in other stars than the Sun, and c) search for more direct supergranulation indicators in stellar spectra.

We also note that targetting stars which are in the low phase of their activity cycle, in order to minimize the impact of activity signals (as proposed for example by \citet{2022MNRAS.514.2259S}), may inadvertently result in a stronger supergranulation signal. Since this signal cannot be mitigated with current methods, that strategy may not be optimal.

\section*{Acknowledgements}

The authors thank Arvind F. Gupta for fruitful discussions. This publication is part of a project that has received funding from the European Research Council (ERC) under the European Union’s Horizon 2020 research and innovation program (Grant agreement No. 865624). NKOS thanks the LSST-DA Data Science Fellowship Program, which is funded by LSST-DA, the Brinson Foundation, the WoodNext Foundation, and the Research Corporation for Science Advancement Foundation; her participation in the program has benefited this work. AM and BL acknowledge funding from a UKRI Future Leader Fellowship, grant number MR/X033244/1. AM acknowledges funding from an STFC small grant, reference: ST/Y002334/1. XD acknowledges the support from the European Research Council (ERC) under the European Union’s Horizon 2020 research and innovation programme (grant agreement SCORE No 851555) and from the Swiss National Science Foundation under the grant SPECTRE (No 200021\textunderscore215200). This work has been carried out within the framework of the NCCR PlanetS supported by the Swiss National Science Foundation under grants 51NF40\textunderscore182901 and 51NF40\textunderscore205606.  SS acknowledges support from the “Programme National de Physique Stellaire“ (PNPS) and “Programme National de Planétologie“ (PNP) of CNRS/INSU co-funded by CEA and CNES. ACC acknowledges support from STFC consolidated grant number ST/V000861/1, and UKRI/ERC Synergy Grant EP/Z000181/1 (REVEAL). This publication makes use of The Data \& Analysis Center for Exoplanets (DACE), which is a facility based at the University of Geneva (CH) dedicated to extrasolar planets data visualisation, exchange and analysis. DACE is a platform of the Swiss National Centre of Competence in Research (NCCR) PlanetS, federating the Swiss expertise in Exoplanet research. The DACE platform is available at \url{https://dace.unige.ch}. Based on observations made with the Italian Telescopio Nazionale Galileo (TNG) operated on the island of La Palma by the Fundación Galileo Galilei of the INAF (Istituto Nazionale di Astrofisica) at the Spanish Observatorio del Roque de los Muchachos of the Instituto de Astrofisica de Canarias. This work made use of \texttt{numpy} \citep[][]{numpy}, \texttt{matplotlib} \citep[][]{matplotlib}, and \texttt{pandas} \citep{pandas} libraries. This work made use of Astropy:\footnote{\url{http://www.astropy.org}} a community-developed core Python package and an ecosystem of tools and resources for astronomy \citep{astropy1, astropy2,astropy3}.

\section*{Data Availability}

This work makes use of the HARPS-N solar RVs, which will be
described and made available in Dumusque et al., submitted. The SDO/HMI images are publicly available at \url{https://sdo.gsfc.nasa.gov/data/}



\bibliographystyle{mnras}
\bibliography{main_text} 




\appendix

\section{Simulations with reduced coverage}
\label{app: reduced}

To test whether the fact that our solar RV time series consist of sequences of 5.3 hours of continuous observations, compared to the day-long timescale of supergranulation, has an effect on the parameters we derived we simulate a month-long data set with observations throughout the whole day. We simulated a granulation, supergranulation and white noise signals, and modelled them with the GP method. We subsample this data set down to  18h/day, 12h/day, and 6h/day and model each subsampled data set. The corner plots of these tests are shown in Figure \ref{fig:reduced sampling}.

These results show that the accuracy of the parameters are not degraded with decreased daily sampling.  We do note, however, that the precision degrades with decreasing sampling. There also seems to be some systematic increase in the deviations of the peak values of the parameter distribution compared to the true values, especially for the granulation parameters, however, these are always within the precision of the measurements.

\begin{figure*}
    \centering
    \includegraphics[width=0.98\textwidth]{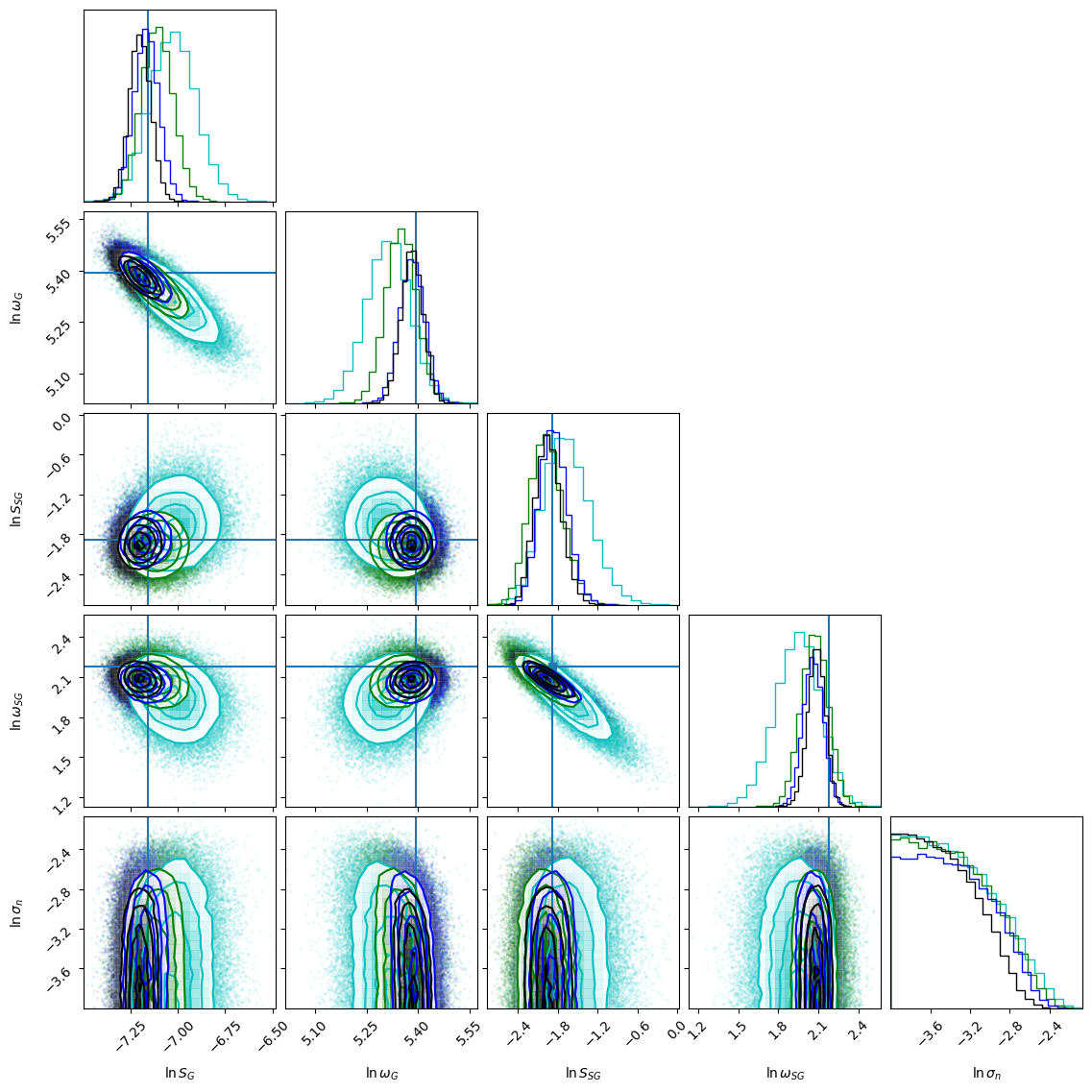}
    \caption{MCMC posterior distribution plots for simulated data with a baseline of 28 days. Observations were simulated for 24 hours (black), 18 hours (blue), 12 hours (green), and 6 hours(cyan) in a 24 hour period. The 1-D posterior distributions for each parameter, marginalised over all the other parameters, are shown by the histograms in the diagonal panels,  with the true parameter values indicated by the blue lines. The 2-D posteriors are shown in the off-diagonal panels, with 1-, 2- and 3-$\sigma$ contours.}
    \label{fig:reduced sampling}
\end{figure*}

\section{Simulations with Activity}
\label{app: resids}

Here we show that activity residuals do not affect the convective signal that we observe. We simulated a 1-year long time-series including activity residuals  (using 1 year’s worth of HARPS-N timestamps and the parameters measured from the observed RVs for that yearly chunk). The activity residuals were simulated using a quasi-periodic celerite SHO term with Q > 1 at the rotation period. The activity signal included was at a period of 13.5 days, to correspond with the activity residuals observed in the periodogram seen in Figure \ref{fig:all data psd}. We then modelled the data with a GP consisting of two SHO aperiodic kernels and a white noise term (with a prior of $[-10;10]$ {m/s}. Figure \ref{fig:corner with activity} shows the result. We see that the inclusion of the activity signal in our simulation does not change the accuracy of the granulation and supergranulation signals found, and the uncertainties are unchanged compared to the case where we do not inject activity residuals. We therefore do not expect these residuals to have a strong effect on the super-granulation properties measured using our GP models. Our explanation for this is that the GP model containing only SG and granulation terms has enough flexibility to absorb longer-term variations, but they do not dominate the variance or characteristic timescale estimates. We note that the white noise term is badly constrained, this is most likely the effect of the activity signal not being modelled.

\begin{figure*}
    \centering
    \includegraphics[width=0.98\textwidth]{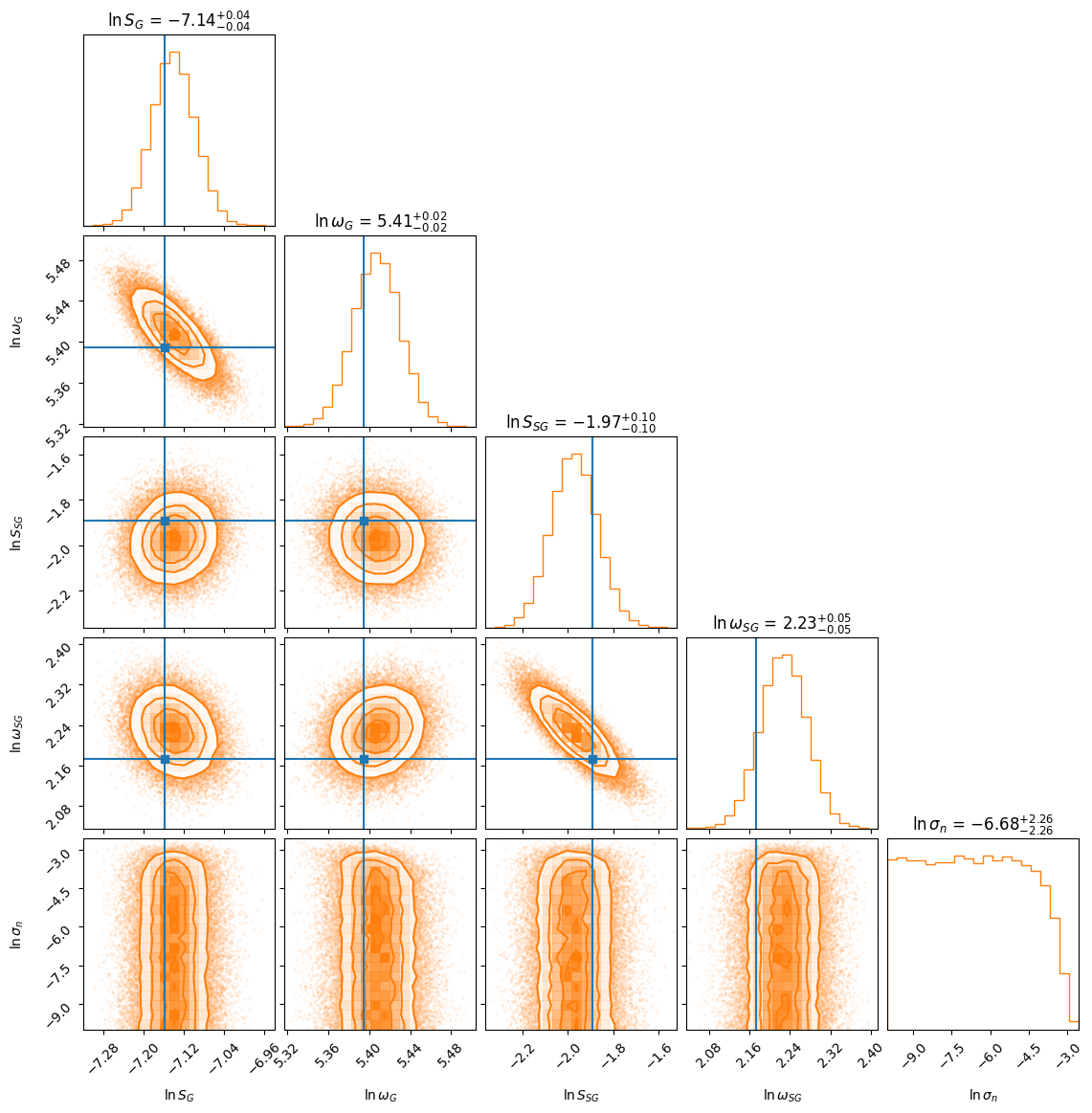}
    \caption{MCMC posterior distribution plots for the the granulation and supergranulation components. The data modelled was simulated with the addition of an activity signal with period = 13.5 hours.  The 1-D posterior distributions for each parameter, marginalised over all the other parameters, are shown by the histograms in the diagonal panels,  with the true parameter values indicated by the blue lines. The 2-D posteriors are shown in the off-diagonal panels, with 1-, 2- and 3-$\sigma$ contours.}
    \label{fig:corner with activity}
\end{figure*}

\section{Corner Plots}
\label{app: corner plots }

Here, we show the corner plots for three chunks representative of the solar cycle for both the SDO (Figure \ref{fig:sdo corner}) and YARARA (Figure \ref{fig:yarara corner}) quiet Sun RVs. We see that in both cases, the results for the three chunks are the same for granulation, while for supergranulation, the chunk corresponding to solar minimum shows diverging results compared to the end of cycle 24 and the start of cycle 25, as is expected. 

\begin{figure*}
    \centering
    \includegraphics[width=0.98\textwidth]{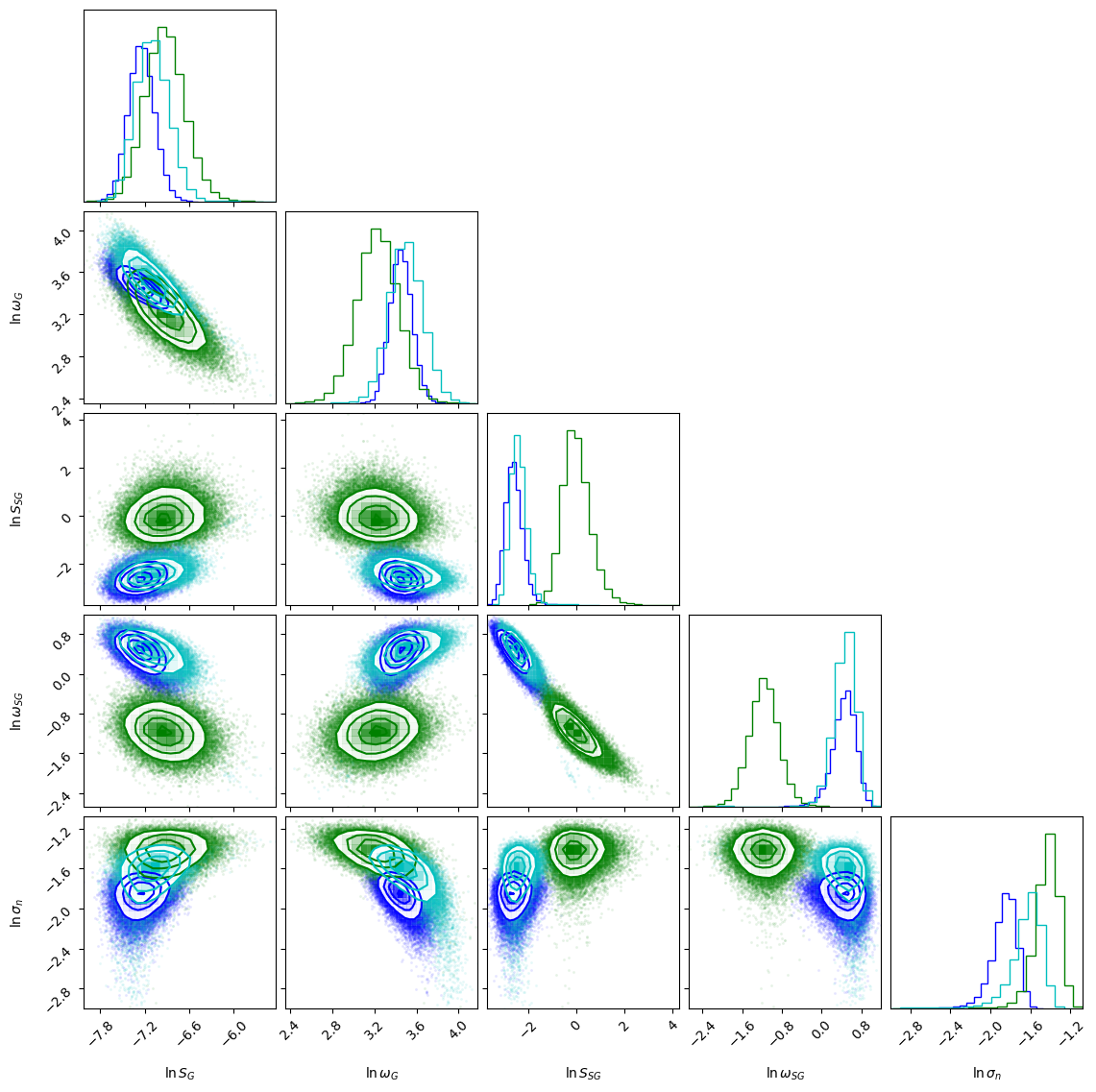}
    \caption{MCMC posterior distribution plots for SDO quiet sun results for 3 sample chunks. The results for the chunk 2016.47 - 2016.55 are in blue, 2018.77-2018.84 in green, and 2022.45-2022.52 in cyan.  The 1-D posterior distributions for each parameter, marginalised over all the other parameters, are shown by the histograms in the diagonal panels. The 2-D posteriors are shown in the off-diagonal panels, with 1-, 2- and 3-$\sigma$ contours.}
    \label{fig:sdo corner}
\end{figure*}

\begin{figure*}
    \centering
    \includegraphics[width=0.98\textwidth]{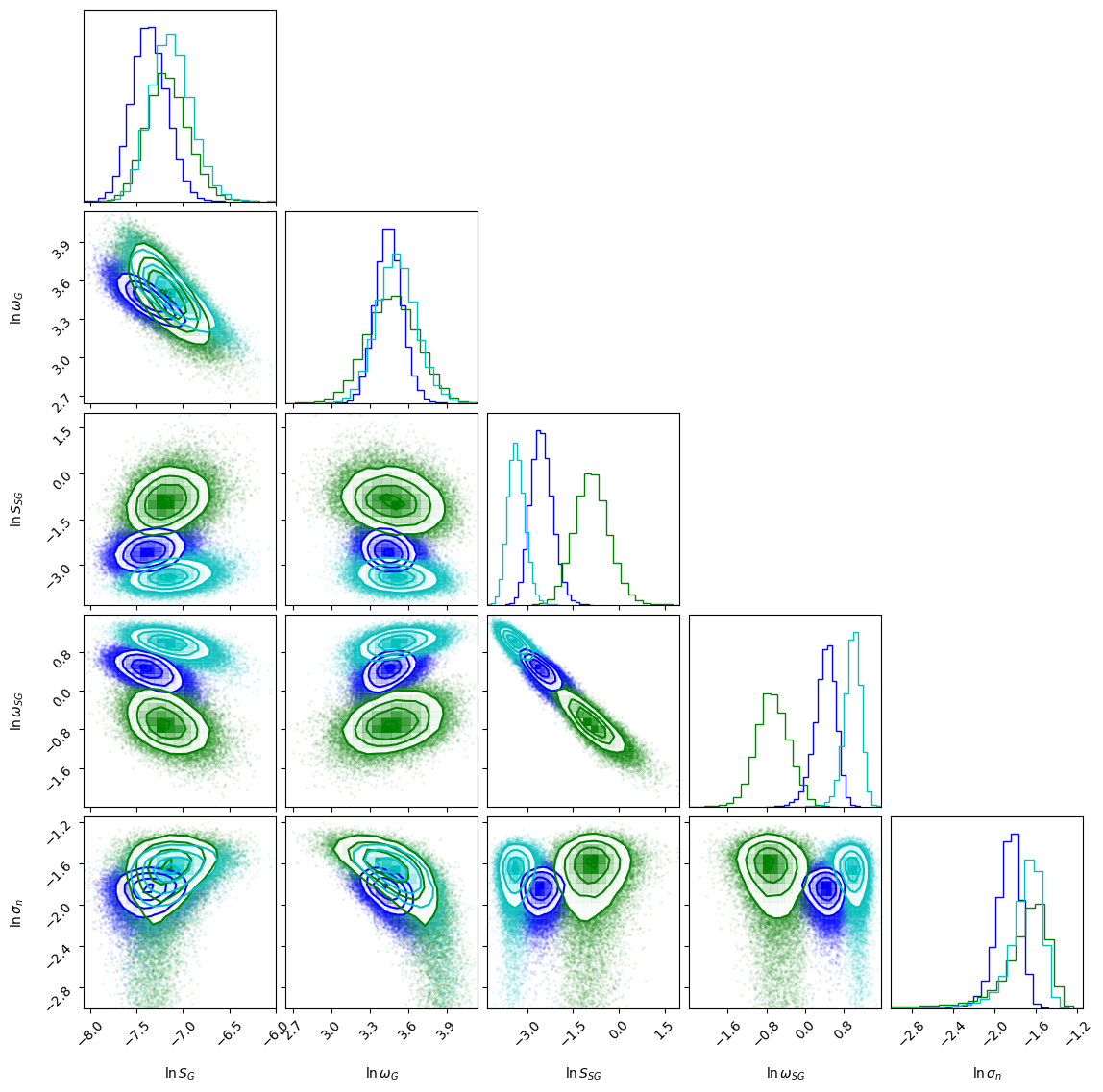}
    \caption{MCMC posterior distribution plots for YARARA quiet sun results for 3 sample chunks. The results for the chunk 2016.47 - 2016.55 are in blue, 2018.77-2018.84 in green, and 2022.45-2022.52 in cyan.    The 1-D posterior distributions for each parameter, marginalised over all the other parameters, are shown by the histograms in the diagonal panels. The 2-D posteriors are shown in the off-diagonal panels, with 1-, 2- and 3-$\sigma$ contours.}
    \label{fig:yarara corner}
\end{figure*}

\section{Correlation plots for 4 week and 1 year long chunks}
\label{p_value appen}

Here we present the correlation plots, like the one shown in Figure \ref{fig:p values 3}, and the corresponding p-values, for the 4 week and 1 year long chunks. 

\begin{figure*}
    \centering
    \includegraphics[width=0.98\textwidth]{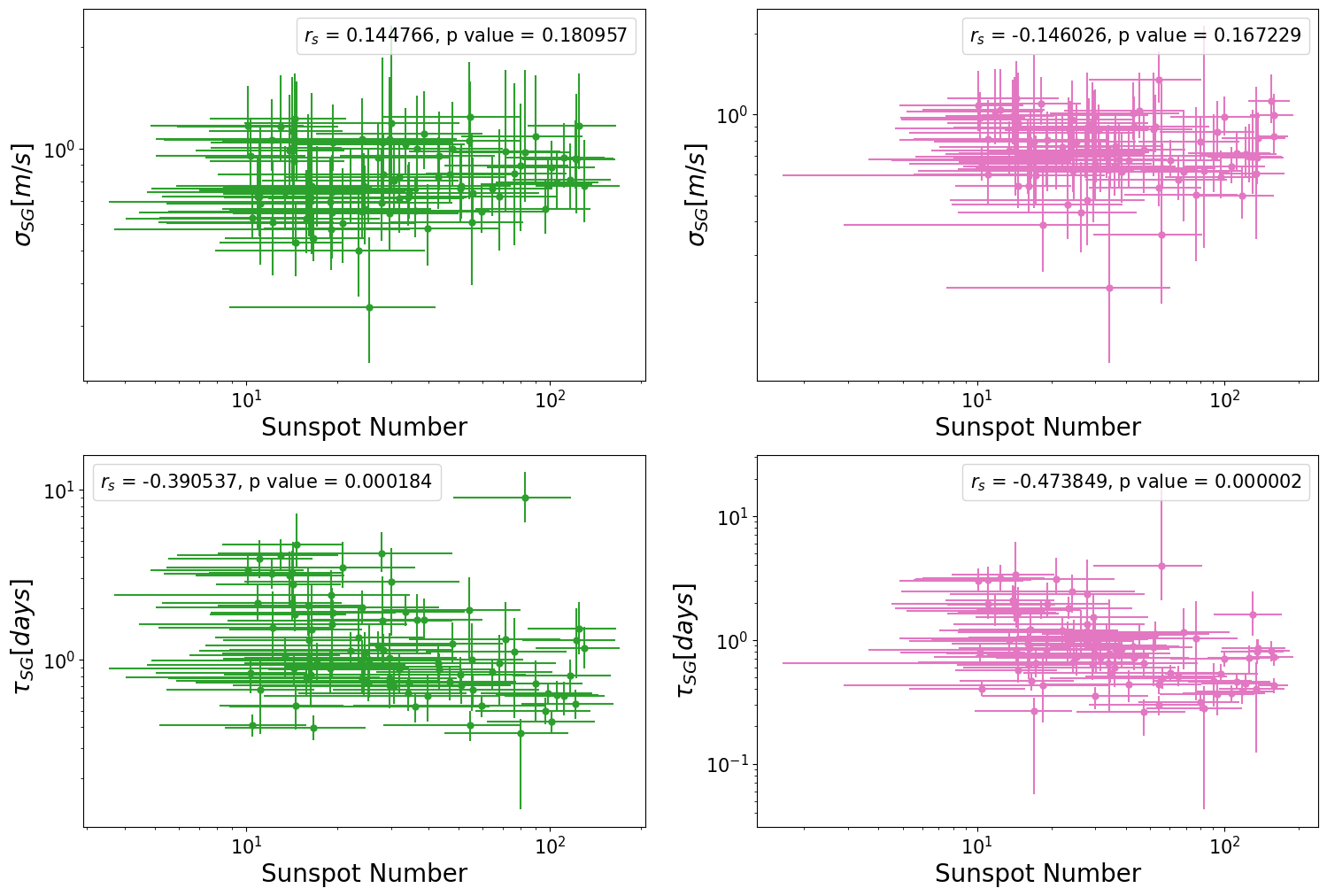}
    \caption{Correlation between the supergranulation parameters obtained from the 4-week  chunks and the Sun-spot numbers, for the SDO-corrected RVs (left, green) and the YARARA-corrected RVs (right, pink). In each panel we also report the $p$-values derived from a Spearman's rank correlation analysis, which correspond to the probability of a similarly correlated dataset arising from white Gaussian noise only.}
    \label{fig:p values 4}
\end{figure*}

\begin{figure*}
    \centering
    \includegraphics[width=0.98\textwidth]{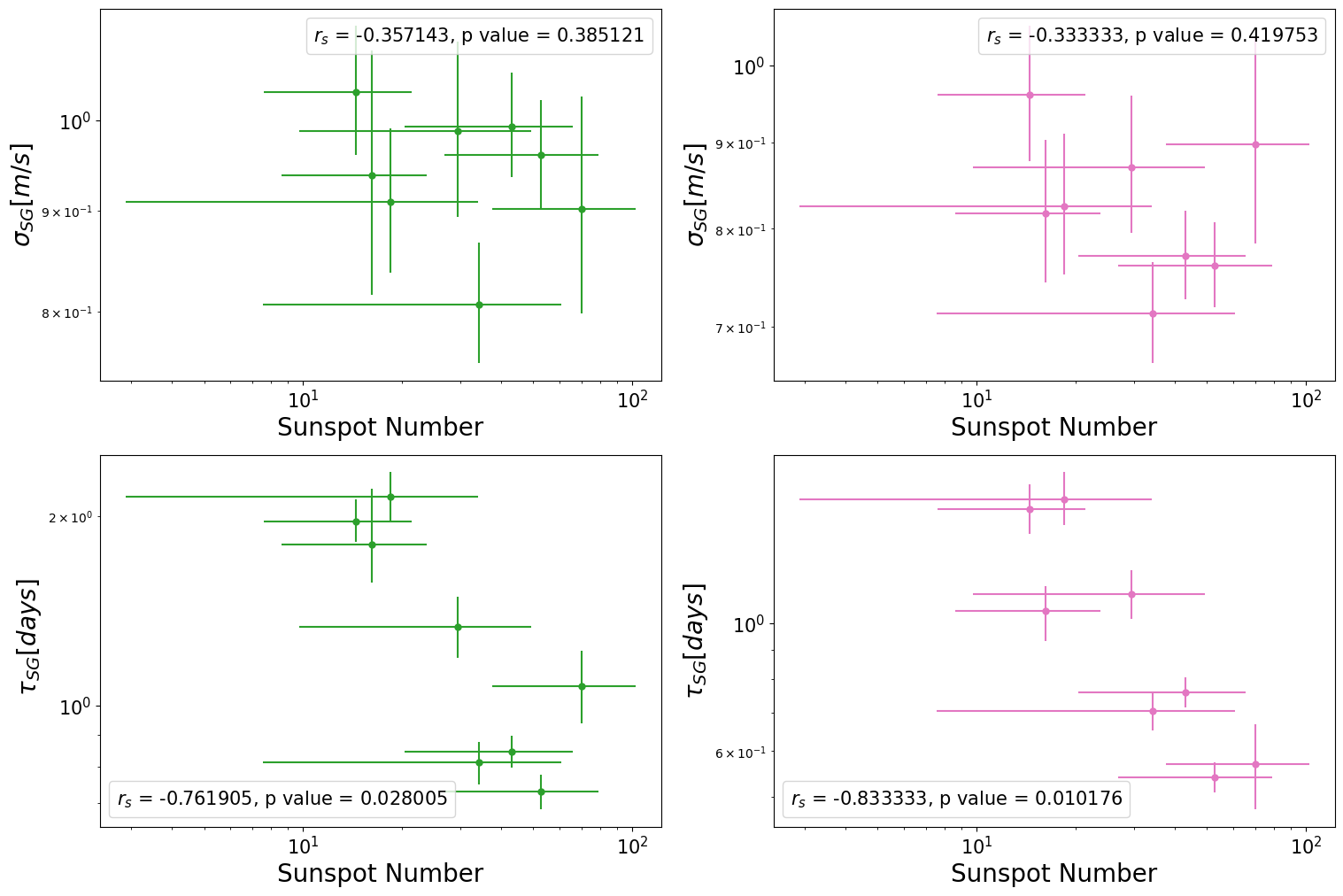}
    \caption{Correlation between the supergranulation parameters obtained from the year long chunks and the Sun-spot numbers, for the SDO-corrected RVs (left, green) and the YARARA-corrected RVs (right, pink). In each panel we also report the $p$-values derived from a Spearman's rank correlation analysis, which correspond to the probability of a similarly correlated dataset arising from white Gaussian noise only.}
    \label{fig:p values 1}
\end{figure*}


\bsp	
\label{lastpage}
\end{document}